\documentclass{article}
\usepackage[utf8]{inputenc}
\usepackage{amsmath}
\usepackage{amsfonts}
\usepackage{amssymb}
\usepackage{graphicx}
\usepackage{color}
\usepackage[left=2.00cm, right=2.00cm, top=2.00cm, bottom=2.00cm]{geometry}
\usepackage{booktabs}
\usepackage{caption}
\usepackage{hyperref}
\usepackage{mathtools}
\usepackage{authblk}
\usepackage{comment}
\usepackage{float}
\restylefloat{table}
\usepackage[toc,page]{appendix}
\usepackage[backend=biber,
style=numeric-comp,natbib=true,maxbibnames=9,
sorting=none,url=false,date=year,doi=false,firstinits=true,isbn=false,bibwarn=true]{biblatex}
\renewbibmacro{in:}{}

\newcommand{\Ghf}{Gauss hypergeometric function~}
\newcommand{\ghf}{general hypergeometric function~}
\newcommand{\f}{$_2 F_1$}
\newcommand{\hf}{hypergeometric function~}
\newcommand{\hfs}{hypergeometric functions~}
\newcommand{\e}{\epsilon}
\newcommand{\eeqref}[1]{Eq. \eqref{#1}}

\newcommand{\mhfs}{multivariable hypergeometric functions~}

\title{$\epsilon$-Expansion of Multivariable Hypergeometric Functions Appearing in Feynman Integral Calculus }
\author[]{Souvik Bera}
\affil{Centre for High Energy Physics, Indian Institute of Science,\\ Bangalore-560012, Karnataka, India}
\affil[]{Email : \texttt {souvikbera@iisc.ac.in}}

\date{}

\begin{document}
	
	\maketitle	
	\begin{abstract}
		We present a new methodology, suitable for implementation on computer, to perform the $\epsilon$-expansion of hypergeometric functions with linear $\epsilon$ dependent Pochhammer parameters in any number of variables. Our approach allows one to perform Taylor as well as Laurent series expansion of multivariable hypergeometric functions. Each of the coefficients of $\epsilon$ in the series expansion is expressed as a linear combination of multivariable hypergeometric functions with the same domain of convergence as that of the original hypergeometric function. We present illustrative examples of hypergeometric functions in one, two and three variables which are typical of Feynman integral calculus.
	\end{abstract}

\vspace{1cm}
\textbf{Keywords} : $\e$-expansion, Feynman integrals, \mhfs
\vspace{1cm}

\section{Introduction}
In the present work we are concerned with the solution to the problem of finding series expansion in the dimensional regularization parameter $\epsilon$ of Feynman integrals originating in the perturbative calculations of quantum field theory.
These integrals can be expressed as multi-variable hypergeometric functions (MHFs) \cite{Bateman:1953,Slater:1966,Exton:1976,Srivastava:1985}.  To give a few examples, two loop massive sunset diagram \cite{Berends:1993ee,Ananthanarayan:2019icl},  three loop vacuum diagram \cite{Gu:2020ypr}, one loop two, three and four point scalar functions \cite{Davydychev:1990cq,Fleischer:2003rm,Boos:1990rg,Tarasov:2000sf,Feng:hypergeometry,Feng:2018zxf,Phan:2018cnz,Davydychev:2005nf} are evaluated in terms of MHFs. Functional relations are used to find one loop Feynman integrals (see \cite{Tarasov:2022clb} and references within) in terms of MHFs. Feynman integrals can also be realized as GKZ hypergeometric system \cite{Kalmykov:2012rr,delaCruz:2019skx,Klausen:2019hrg,Feng:2019bdx,Ananthanarayan:2022ntm}. In dimensional regularization \cite{tHooft:1972tcz}, the space-time dimension $D$ (more precisely, integer multiples of $D/2$) of the Feynman integral appears linearly in the Pochhammer parameters of the corresponding MHFs and the ratios of the scales involved in the integral take its place in the arguments of MHFs. Usually the result of Feynman integral is expressed as a series in the parameter $\epsilon = (4-D)/2$. Thus, it is important to find efficient algorithms to determine the coefficients of  the $\e$-expansion of the MHFs analytically as well as numerically that appears in the Feynman integral calculus. There are computer programs \cite{Borowka:2017idc,Smirnov:2021rhf}, based on sector decomposition method, that can perform the series expansion of Feynman integrals numerically.

The $\e$-expansion of the one variable \Ghf $_2F_1$ \cite{Kalmykov:2006pu} and $_pF_{p-1}$ \cite{Kalmykov:2007pf,Kalmykov:2008ge,Greynat:genhypfun} are well studied. There are publicly available packages that can perform the expansion of $_{p}F_{p-1}$ around integer and half-integer Pochhammer parameters analytically \cite{Moch:Xsummer,HypExp,HypExp2} and numerically \cite{NumExp}. The $\e$-expansion of double variable Appell  and Kamp\'e de F\'eriet functions are discussed in \cite{Greynat:appellkdf,Greynat:2014jsa,Moch:2001zr,Weinzierl:halfint,Weinzierl:code}. In \cite{Yost:2011wk,Bytev:2012ud,Kalmykov:2020cqz}, differential equation approach is used to find the series expansion of MHFs related to Feynman integrals (see also \cite{Blumlein:2021hbq,Bluemlein:2022eym}). 
One can find a comprehensive review of \mhfs and their $\e$-expansion related to the Feynman integral calculus in \cite{Kalmykov:2020cqz}. Despite all this work, it may be recognized that, a general procedure to find the series expansion of any MHF in $\e$ does not exist in the literature.

In view of the above consideration, we present an efficient algorithm, that can be implemented on the computer, to evaluate analytically the series expansion of hypergeometric functions with any number of variables in the parameter $\e$ that appears linearly in the Pochhammer parameters. There is no restriction on other parameters of the function and these could be any complex numbers. In contrast to the available methods, each of the coefficients of the $\e$ expansion is expressed in terms of MHFs. The domain of convergence of the \hfs appearing in the coefficients of the $\e$-expansion is the same as that of the starting hypergeometric function. Thus, in the context of Feynman integrals,  the hypergeometric structure remains intact throughout this method. 

In general
the expansion coefficients are expressed in terms of Harmonic polylogarithms (HPLs) \cite{Remiddi:1999ew} and nested sums \cite{Vermaseren:1998uu,Blumlein:2003gb,Davydychev:2003mv}. It is observed that not all MHFs can be expanded in terms of HPLs and nested sums, which leads to further generalization, functions with elliptic integrals. On the other hand, MHFs provide a function space for expressing not only Feynman integrals but also the coefficients of its $\e$-expansion. The summation indices of MHFs, in principle, range over all non-negative integers, but for practical calculation, any large finite number (N) can be taken as the upper limit. The accuracy of the result will depend on N. One has to choose a suitable upper limit N to make sure that the summation is converged and the desired accuracy is reached. Thus, the numerical evaluation of MHFs is not a difficult task.  Inside the associated domain of convergence, these analytic expressions can be used to find numerical values. We have applied our methodology to one, two and three variable hypergeometric functions and results are found to be consistent.

The article is organized as follows. In Section \ref{sec:Methodology} we explain each of the steps of our methodology in details. A number of illustrative examples of \hfs   are presented in Section \ref{sec:examples} which includes one variable Gauss $_2F_1$, $_2F_2$, double variable Appell $F_1$ and $F_2$ functions and three variable Lauricella function $F_C^{(3)}$. The provided examples are of Appell  $F_1$ and Lauricella $F_C^{(3)}$, which are motivated by one loop three point function and two loop sunset integral respectively. The example section is  followed by the summary. 

We note here that the ancillary notebook contains the first few terms, including those presented in the article, of the $\e$-expansions of the hypergeometric functions that are considered in Section \ref{sec:examples}.

	\section{The Methodology}\label{sec:Methodology}
	
	The presented approach of finding the series expansion is applicable when the parameter $\e$ appears linearly in the Pochhammer parameters of a MHF, which is usually the case when one expresses a Feynman integral in terms of MHF.
	
	Our method employs two well-established techniques from the literature. The first technique deals with the derivative of a MHF with respect to its Pochhammer parameters. We follow a similar approach introduced by Anacarani-Gasaneo in \cite{Ancarani1F1, Ancarani2009, Ancarani2010} for one variable  hypergeometric functions and its generalizations for more than one variables in \cite{Ancarani2017, BYTEV2020114911}, which expresses the result of the Pochhammer derivative of any MHF as  MHFs with one extra summation fold, but having the same set of variables. The second technique concerns about the differential operators that generate the contiguous relations of a MHF \cite{Takayama1989}. We apply these techniques  to perform the task of the series expansion of a MHF in a systematic way.
	
	This method can be divided into five steps.
	
	\begin{itemize}
		\item \textit{Step 1}: Determine whether the expansion of the given MHF (say, $F(\e)$) is of Taylor or Laurent type simply by observing the lower Pochhammer parameters (i.e., the Pochhammer parameters that appear in the denominator of the series representation of the MHF)
		\item \textit{Step 2}: If the series expansion of the given MHF is of Taylor type, then find the series expansion from the definition of the Taylor expansion
		\begin{align}
			F(\e) = \sum_{i=0}^\infty \frac{\e^i}{i!} \frac{d^i}{d \e^i} F(\e)\Big\vert_{\e=0}
		\end{align}
		\item \textit{Step 3}: If the series expansion of the given MHF (primary function) is of Laurent type, then find a secondary MHF (say, $G(\e)$) and a differential operator (say, $H$) such that,  the secondary MHF can be expanded in Taylor series and the differential operator can be used to relate the secondary MHF with the primary MHF
		\begin{align}
			F(\e) =  H \bullet G(\e)
		\end{align}
		Calculate the Taylor expansion of the secondary MHF following \textit{step 2}
			\begin{align}
			G(\e) = \sum_{j=0}^\infty \e^j G_j
		\end{align}
		\item \textit{Step 4}: Find the corresponding differential operator (i.e., $H$)
		
		\item \textit{Step 5}: Expand the differential operator in series and apply it on the Taylor expansion of the secondary MHF and collect the coefficients of different powers of $\e$
		\begin{align*}
			F(\e) &=  H \bullet G(\e)\\
			&= \left[ \sum_{i=-n}^\infty \e^i H_i \right] \bullet \left[ \sum_{j=0}^\infty \e^j G_j \right]
		\end{align*}
	\end{itemize}
	
	We now elaborate each of the steps in details.

	\subsection{\textit{Step 1}: Determination of the type of the series expansion}

	
	The expansion of a MHF around $\e=0$ can either be of Taylor type of Laurent type. If the lower Pochhammer parameters (i.e., Pochhammer parameters in the denominator) of a MHF  has the form
	\begin{align}
		(B_0+B_1 \e)_*
	\end{align}
	where $B_0$ is zero or negative integer, then it is called \textit{singular}. The ``*'' denotes linear combination of summation indices which does not play any role in what follows. If one or more lower Pochhammer parameters of MHF are singular, then $\e$-expansion of that function may be of Laurent type. If the lower Pochhammer parameters are non-singular, the $\e$-expansion of the corresponding function is of Taylor type. We discuss each of the situations below.
	

	\subsection{\textit{Step 2}: Taylor expansion of MHFs}\label{sec:TaylorSeries}
		The Taylor series expansion around $\e=0$ of a MHF, which we denote by $F(\e)$, suppressing dependencies of all the other parameters but $\e$, can be found by taking successive derivatives with respect to $\e$
	\begin{align}\label{eqn:TaylorExp}
		F(\e) = \sum_{i=0}^\infty \frac{\e^i}{i!} \frac{d^i}{d \e^i} F(\e)\Big\vert_{\e=0}
	\end{align}
	 In  practice we only need first few terms of the series expansion.

	Since the parameter $\e$ takes its place in the Pochhammer parameters of a MHF, taking derivative of Pochhammer parameters with respect to its argument plays crucial role in finding the Taylor series expansion. Derivatives of Gauss hypergeometric function $_2F_1$  with respect to the Pochhammer parameters are studied in \cite{Ancarani2009}. Similar study for general  $_p F_q$ function can be found in \cite{Ancarani2010,Kang2015,Fejzullahu2017,Greynat:genhypfun}. Discussions on the Pochhammer derivative of hypergeometric functions of more than one variable can be found in \cite{Sahai2009,Ancarani2017,Greynat:appellkdf,Greynat:2014jsa,BYTEV2020114911}. In this section we show how the Pochhammer derivative of a hypergeometric function with $n$ fold summations can be represented as  \hf with $n+1$ fold summations and the same set of variables.  This fact is used to find the Taylor series expansion of MHF in $\e$, whose coefficients  themselves are MHFs.
	
	\subsubsection{Derivative of Pochhammer parameters}
	We start by taking derivative of Pochhammer parameter with respect to its argument. In what follows, we assume that $m,n,k$ take non-negative integer values.
	\begin{align}\label{eqn:Pochderi}
		\frac{d (a)_n}{d a} = (a)_n \left(\psi ^{(0)}(a+n)-\psi ^{(0)}(a)  \right) = (a)_{n} \sum_{k=0}^{n-1} \frac{1}{a+k}=(a)_{n} \frac{1}{a} \sum_{k=0}^{n-1} \frac{(a)_{k}}{(a+1)_{k}}
	\end{align}
	where $\psi^0$ is polygamma function. If $a$ is $\e$ dependent, the chain rule can be used to find the derivative with respect to $\e$,
	\begin{align}
		\frac{d (a)_m}{d \e} = \frac{d a}{d \e}\frac{d (a)_m}{d a} 
	\end{align}
	
	Derivative of Pochhammer in other circumstances can be obtained by Pochhammer identities (listed in the Appendix \ref{Appendix:Pochid}), chain rule and \eeqref{eqn:Pochderi}. We list few cases below.
	
	\begin{itemize}
		\item Derivative of powers of Pochhammer  can be found by applying chain rule first and then \eeqref{eqn:Pochderi}. For example,
		\begin{align}
			\frac{d}{da} \left[(a)_m\right]^p = p \left[(a)_m\right]^{p-1} \frac{d}{da} (a)_m
		\end{align}
		\item To find the derivative of $(a)_{p m}$, where $p$ is a positive integer, Gauss's multiplication formula can be used to simplify $(a)_{p m}$ and then chain rule and \eeqref{eqn:Pochderi} can be applied. For example,
		\begin{align}
			\frac{d (a)_{2m}}{d a} = \frac{d}{d a} \left[2^{2 m} \left(\frac{a}{2}+\frac{1}{2}\right)_m \left(\frac{a}{2}\right)_m\right] = 2^{2 m} \frac{d}{d a}\left[\left(\frac{a}{2}+\frac{1}{2}\right)_m\right] \left(\frac{a}{2}\right)_m + 2^{2 m} \left(\frac{a}{2}+\frac{1}{2}\right)_m \frac{d}{d a} \left[\left(\frac{a}{2}\right)_m\right]
		\end{align}
		\item $ (a)_{-m}$ can be simplified using the Pochhammer identities and then the derivative of the result can be found using \eeqref{eqn:Pochderi}
		\begin{align}
			\frac{d}{d a} (a)_{-m} = \frac{d}{d a} \left[\frac{(-1)^{ m}}{(1-a)_{ m}}\right] =  (-1)^{ m} \frac{d}{d a} \left[\frac{1}{(1-a)_m}\right] 
		\end{align}
		\item Using the Pochhammer identities, Pochhammer parameter like $(a)_{m+n}$ can be written as products of several Pochhammers parameters, Then chain rule and \eeqref{eqn:Pochderi} can be used to find the derivative of $(a)_{m+n}$.
		\begin{align}
			\frac{d}{d a} (a)_{m+n} = \frac{d}{d a} \left[(a)_m (a+m)_n\right] = \frac{d}{da}\left[(a)_m\right]  (a+m)_n + (a)_m \frac{d}{da}\left[(a+m)_n\right]
		\end{align}
	\end{itemize}
	
	Equipped with all the techniques of finding derivative of Pochhammer parameters, we are ready to consider Pochhammer derivative of MHF.
	
	\subsubsection{Pochhammer derivative of MHF} \label{sec:Pochhammerderivative}
	We assume the parameter with respect to which the derivative of a MHF is to be taken is $a$. First the Pochhammer containing $a$ can be simplified using Pochhammer identities. Then the chain rule of derivative can be applied. Essentially, the task of finding the Pochhammer derivative of a MHF boils down to evaluating a number of expressions that look like the following 
	\begin{align}
		\sum_{n=0}^\infty B(n) \frac{x^n}{n!} \frac{d}{da}(a)_n
	\end{align}
	Following \cite{BYTEV2020114911}, the summation of the index $n$ is explicitly shown as the Pochhammer parameter that is being differentiated contains $n$. Here $B(n)$ collectively denotes the summation of indices  other than $n$ of the MHF. Using \eeqref{eqn:Pochderi} we write
	\begin{align}
		\sum_{n=0}^\infty B(n) \frac{x^n}{n!} \frac{d}{da}(a)_n =\sum_{n=0}^\infty B(n) \frac{x^n}{n!} (a)_{n} \frac{1}{a} \sum_{k=0}^{n-1} \frac{(a)_{k}}{(a+1)_{k}}
	\end{align}
	Now shifting $n\rightarrow n+1$,
	\begin{align}
		\sum_{n=0}^\infty B(n) \frac{x^n}{n!} \frac{d}{da}(a)_n =\sum_{n=0}^\infty B(n+1) \frac{x^{n+1}}{(n+1)!} (a)_{n+1} \frac{1}{a} \sum_{k=0}^{n} \frac{(a)_{k}}{(a+1)_{k}}
	\end{align}
	Reshuffling the summation of indices $n,k$ as,
	\begin{align}
		\sum_{n=0}^\infty \sum_{k=0}^n A(n,k) \rightarrow \sum_{n,k=0}^\infty A(n+k,k)
	\end{align}
	we find
	\begin{align}\label{eqn:deriformula}
		\sum_{n=0}^\infty B(n) \frac{x^n}{n!} \frac{d}{da}(a)_n 
		&= \sum_{n,k=0}^\infty B(n+k+1) \frac{x^{n+k+1}}{(n+k+1)!} (a)_{n+k+1} \frac{1}{a} \frac{(a)_{k}}{(a+1)_{k}} \nonumber\\
		&= x \sum_{n,k=0}^\infty B(n+k+1) \frac{(1)_k (1)_n (a)_k (a+1)_{k+n}}{(a+1)_k (2)_{k+n}} \frac{x^n x^k}{k!n!}
	\end{align}
	Hence, taking the Pochhammer derivative of MHF introduces a new summation index. Thus the summation fold is increased by one. This formula (\eeqref{eqn:deriformula}) is used to express each of the coefficients of the Taylor series expansion of a MHF (\eeqref{eqn:TaylorExp}) in terms of higher summation fold MHFs. 
	
	We observe that derivative of a Pochhammer parameter introduces a polygamma function (\eeqref{eqn:Pochderi}), which is then written as a series with new summation index. A MHF that contains polygamma function inside the summand (i.e. the argument of the polygamma function contains summation indices),  has the same domain of convergence as that of MHF with out any polygamma function (see Appendix B of \cite{Friot:convergence}).  Hence the domain of convergence of a MHF obtained by taking Pochhammer derivative of a original MHF, is the same as that of the original MHF. This fact in turn implies that, the domain of convergence of MHFs appearing in each of the coefficients of $\e$ of the Taylor series expansion of original MHF is same  as the domain of convergence of the original MHF.
	
	\subsection{\textit{Step 3}: Construction of the secondary MHF}\label{sec:Laurent}
	
	As discussed above, if one or more lower Pochhammer parameters of the given MHF (primary function) are singular, then $\e$ expansion of that function may be of Laurent type. In such cases, one can find the secondary MHF by replacing the singular lower Pochhammer parameter by
	\begin{align*}
			(B_1 \e +B_0)_* \rightarrow 	(B_1 \e +1)_*
	\end{align*}
	The \hf with non-singular lower Pochhammer parameters (i.e. secondary function) can be expanded in $\e$ as Taylor series as prescribed in Section \ref{sec:TaylorSeries}.

	
	\subsection{\textit{Step 4}: The differential operator}

	There exist contiguous relations that relate MHFs whose Pochhammer parameters differ by integers. An algorithm based on Gr\"{o}bner bases calculation is proposed by Takayama \cite{Takayama1989} to find the differential operators that generates such relations. These operators play a crucial role in the differential reduction technique of MHFs \cite{Hyperdire,Hyperdire2}. For our purpose we only need one kind of operator, called the \textbf{step down} operator, to decrease the argument of the lower Pochhammer parameters by unity. 
	
	Let us define a MHF as
	\begin{align} \label{eqn:mhfdefinition}
		F(\textbf{a},\textbf{b},\textbf{x}) = \sum_{\textbf{m}\in \mathbb{N}_0^r}  \frac{\Gamma(\textbf{a}+ \mu . \textbf{m} )/ \Gamma(\textbf{a}) }{\Gamma(\textbf{b}+ \nu . \textbf{m} )/ \Gamma(\textbf{b}) } \frac{\textbf{x}^\textbf{m}}{\textbf{m}!} = \sum_{\textbf{m}\in \mathbb{N}_0^r}  \frac{(\textbf{a})_{\mu . \textbf{m}} }{(\textbf{b})_{\nu . \textbf{m}} } \frac{\textbf{x}^\textbf{m}}{\textbf{m}!} = \sum_{\textbf{m}\in \mathbb{N}_0^r} A(\textbf{m}) \textbf{x}^\textbf{m}
	\end{align}
	Here $\textbf{a}, \textbf{b}$ and $\textbf{m}$ are vectors of length $p,q$ and $r$ respectively. $\mu$ and $\nu$ are matrices of size $p \times r$ and $q \times r$ respectively with integers as their elements. The expression is the definition of the $r$ variable \hf with parameters $\textbf{a}$ and $\textbf{b}$. $\mathbb{N}_0$ denotes natural numbers including zero. We have used the vector notation here.
	\begin{itemize}
		\item $\textbf{a} := \{a_1 , a_2, \dots, a_n\}$
		\item $\textbf{x}^\textbf{m} := \prod_{i=1}^n x_i^{m_i}$
		\item $\textbf{m}! := \prod_{i=1}^n (m_i!)$
		\item $\Gamma(\textbf{a}) := \prod_{i=1}^n \Gamma(a_i)$
		\item $(\textbf{a})_\textbf{m} :=\prod_{i=1}^n (a_i)_{m_i} =\prod_{i=1}^n \Gamma(a_i+m_i)/\Gamma(a_i) $
		\item $\sum_{\textbf{m}\in \mathbb{N}_0^n} := \sum_{m_1 =0}^\infty \dots \sum_{m_n =0}^\infty$
	\end{itemize}
	
	These MHFs are known to satisfy partial differential equations \cite{Bateman:1953}. Let,
	\begin{align}
		P_i = \frac{A(\textbf{m}+\textbf{e}_\textbf{i})}{A(\textbf{m})} = \frac{g_i(\textbf{m})}{h_i(\textbf{m})} \hspace{1cm}\text{,~} i=1,\dots,r
	\end{align}
	where $\textbf{e}_\textbf{i}$ is unit vector with $1$ in its i-th entry. The annihilators $L_i$ of  $	F(\textbf{a},\textbf{b},\textbf{x})$ (\eeqref{eqn:mhfdefinition}) are given by
	\begin{align}
		L_i = \left[ h_i(\pmb{\theta}) \frac{1}{x_i} - g_i(\pmb{\theta}) \right]
	\end{align}
	where $\pmb{\theta} = \{\theta_1,\dots,\theta_r \}$ is a vector containing  Euler operators $\theta_i= x_i\partial_{x_i}$
	
	Following \cite{Takayama1989,Hyperdire}, we define the unit step down operator for the lower Pochhammer parameter

	\begin{align}
	F(\textbf{a},\textbf{b},\textbf{x}) = 	\frac{1}{b_i}\left(\sum_{j=1}^{r} \nu_{i j} \theta_{x_j} +b_i\right) \bullet	F(\textbf{a},\textbf{b}+\mathbf{e_i},\textbf{x}) = H(b_i) \bullet  F(\textbf{a},\textbf{b}+\mathbf{e_i},\textbf{x})
	\end{align}
	
	Here the bullet means the action of the operator $H$ on a function. The step down operator, if needed, can be applied multiple times to decrease any lower Pochhammer parameter by suitable integer. In such cases, the product of the unit step down operators modulo the ideal generated by $L_i$'s can be taken as the step down operator.
	
	\begin{align}
		H = \left[\prod_{i=1} H(b_i) \right]/ \langle L_1,\dots,L_r\rangle
	\end{align}
	The calculation of Gr\"obner bases of the annihilators  and the reduction of the step down operator with respect to the ideal generated by the annihilators  can be found using computer programs \cite{Chyzak1998} in Maple  as well as \cite{koutschan2010fast,KoutschanPhD} in Mathematica. We present an example of step down operator for the one variable \Ghf below. 
	

	\subsubsection*{Example of \Ghf}
	
	The unit step down operator for the \Ghf is given by $H(c) = \frac{1}{c} (\theta+c)$, i.e.
	\begin{align}
		_2F_1(a,b;c;x) = H(c) \bullet~ _2F_1(a,b;c+1;x) 
	\end{align}
	Here the Euler operator $\theta =  x \partial_x $. The above identity can be proved from the definition of the Gauss hypergeometric function.
	\begin{align}
		c~ _2F_1(a,b;c;x) &= \sum_{m=0}^\infty \frac{c  (a)_m (b)_m}{(c)_m } \frac{x^m}{m!}\nonumber\\
		&= \sum_{m=0}^\infty \frac{(c+m)  (a)_m (b)_m}{(c+1)_m } \frac{x^m}{m!}\nonumber\\
		&= (\theta+ c)\bullet~_2F_1(a,b;c+1;x)
	\end{align}
	In the second step we have used the Pochhammer identity $\frac{c}{(c)_m} =\frac{c (c+m) \Gamma(c)}{(c+m) \Gamma(c+m)} = \frac{(c+m)}{(c+1)_m}$ and the last equality is obtained using $\theta \bullet x^m =  m x^m$.
	
	The unit step down operator can be used repetitively to find a relation between $_2F_1(a,b;-n + c \e;x)$ to $_2F_1(a,b;c \e +1;x)$, where $n$ is any positive integer. Note that the later \hf can be expanded in Taylor series. For example
	\begin{align}
		_2F_1(a,b;c-1;x) =  H(c-1)  H(c) \bullet~ _2F_1(a,b;c+1;x)
	\end{align}
	In the case, the step down operator $H$ is $H(c-1) H(c)$ modulo $L$, where $L$  the annihilator of the hypergeometric function $_2F_1(a,b;c+1,x)$.
	\begin{align}
		L\bullet~_2F_1(a,b;c+1;x) = \left[-a b + \left(-x (a+b+1)+c+1\right) \partial_x -(x-1) x\partial_x^2 \right]\bullet~_2F_1(a,b;c+1;x)=0
	\end{align}
Hence we find the step down operator
	\begin{align}
		H =  \left(1-\frac{a b x}{(c-1) c (x-1)}\right) -\frac{x (a+b+1)-2 c x+c-1}{(c-1) c (x-1)} \theta 
	\end{align}
	which matches with the result obtained from the package \texttt{pfq.m} \cite{Hyperdire}. The above differential operator when  made to act on $_2F_1(a,b;c+1;x)$ yields
	\begin{align}
		_2F_1(a,b;c-1;x) &= \left(1-\frac{a b x}{(c-1) c (x-1)}\right) \, _2F_1(a,b;c+1;x)\nonumber\\
		&+\frac{a b x (x (-(a+b+1))+c (2 x-1)+1) }{(c-1) c (c+1) (x-1)}   \, _2F_1(a+1,b+1;c+2;x)
	\end{align}

\subsection{\textit{Step 5}: Argument derivative of MHF}

Usually, the step down operators contain derivative operators with respect to the arguments of the corresponding MHF (see the examples in  Section \ref{sec:examples}). We call the action of these derivative operators on MHF as `argument derivative' in order to distinguish it from Pochhammer derivative discussed above.

Similar to the discussion in Section \ref{sec:Pochhammerderivative}, we explicitly write the summation of the index $n$ related to the argument with respect to which the differential is being taken.
\begin{align}
\frac{d}{dx}	\sum_{n=0}^\infty B(n) \frac{x^n}{n!}
\end{align}
Here, as before, $B(n)$ collectively denotes the dependence of all the other summation indices in a MHF. 

So we have
\begin{align}
\frac{d}{dx}	\sum_{n=0}^\infty B(n) \frac{x^n}{n!} = 	\sum_{n=1}^\infty B(n) \frac{x^{n-1}}{(n-1)!} = 	\sum_{n=0}^\infty B(n+1) \frac{x^n}{n!}
\end{align} 
We have shifted $n\rightarrow n+1$ to get the last equality. Hence the argument derivative of a MHF is also  another MHF. It is straightforward to realize that the domain of convergence of the resultant MHF remains same as the initial one. 

Let us summarize the whole procedure. For a given MHF, we first check if any of its lower Pochhammer parameters is singular. If all the lower Pochhammer parameters are non-singular, then the expansion of that MHF is of Taylor type and can be computed by taking successive derivatives with respect to $\e$ as discussed in Section \ref{sec:TaylorSeries}. The coefficients of each order of $\e$ are expressed in terms of MHFs with the same domain of convergence as that of the original MHF. If one or more lower Pochhammer parameters of the given MHF are singular, then the expansion of the given MHF may be of Laurent type. In that case, a suitable step down operator is found that relates the given MHF (primary function) with a secondary MHF, which can be expanded in Taylor series. Once the Taylor series expansion of the secondary MHF and the step down operator are found, the Laurent series expansion of the primary MHF can be found by applying the stepdown operator on the Taylor expansion of the secondary MHF. As in the case of Taylor series expansion, each of the coefficient of the Laurent series expansion are expressed in terms of MHFs and the domain of convergences remains same as the given MHF.

It is worth pointing out that, one has to obtain suitable analytic continuation before proceeding to find the $\e$-expansion. Finding analytic continuations of general MHF is an active field of research.
There are many results of analytic continuations of MHFs available in the mathematics literature \cite{Bateman:1953,Slater:1966,Exton:1976,Srivastava:1985,Bezrodnykh:Horn3var,Bezrodnykh:Horn_arbitrary_var,Bezrodnykh:LauricellaFD,Bezrodnykh:LauricellaFDz_z,Bezrodnykh:Lauricellafunctions}. Recently the analytic continuations of the double variable Appell $F_2$ are found in \cite{AppellF2} which are the backbone of the numerical package \texttt{AppellF2.wl}. One can find the numerical value of the Appell $F_2$ function for general values of Pochhammer parameters and real values of $x$ and $y$ using this package. Some new analytic continuations of the Appell $F_4$ \cite{Ananthanarayan:F4}, Horn $H_1, H_5$ functions \cite{Bera:H1H5} are found. Analytic continuations of three variable Srivastava $H_C$ functions are obtained in \cite{Friot:2022dme}. Study of analytic continuations and numerical evaluations of other \hfs is ongoing \cite{ASST,ASS}. 
Recently, the process of finding analytic continuations of MHFs is automated with Mathematica based \texttt{Olsson.wl} package \cite{Olssonwl}. This package  eases the calculation of finding certain analytic continuations of MHFs associated with a Feynman integral. Once a suitable analytic continuation is obtained, the series expansion in $\e$  of the resultant MHFs can be performed.

It is to be noted that, the $\e$-expansion of a MHF at its singular points have to treated carefully. For example, the Gauss ${}_2F_1(1,1;2+\epsilon;z)$ can be expanded around $z=1$ in the series of $\e$ as
\begin{align*}
	{}_2F_1(1,1;2+\epsilon;z) = -\frac{1+\epsilon}{z} \ln(1-z) + O(\epsilon^2)
\end{align*}
This series expansion is valid around $z=1$ but certainly not at $z=1$. Starting from the defining series representation of the Gauss $_2F_1$ or its analytic continuation around $z=1$, we can find the same series expansion as above.  But at $z=1$, using the Gauss summation theorem,
\begin{align*}
	_2F_1 (a,b;c;1) = \frac{\Gamma(c)\Gamma(c-a-b)}{\Gamma(c-a)\Gamma(c-b)},\hspace{1cm} \Re(c)> \Re(a+b)
\end{align*}
 one can find the series expansion as
\begin{align*}
	{}_2F_1(1,1;2+\epsilon;1) = \frac{1+\epsilon}{\epsilon}
\end{align*}
Thus, a careful analysis is required for the series expansion of a MHF at its singular points, which is beyond the scope of the presented method.

We evaluate the $\e$-expansion of some MHFs to demonstrate the methodology below.

	\section{Examples}\label{sec:examples}
	
	\subsection{Gauss $_2F_1$ example with half integer Pochhammer}\label{sec:Gausshalfinteger}
	We begin this section by providing the $\e$-expansion of the following Gauss hypergeometric function
	\begin{align}
		G \vcentcolon=~ _2F_1(3,2;\e-3/2;x) = \sum_{m=0}^\infty \frac{(2)_m (3)_m }{ \left(\e-\frac{3}{2}\right)_m} \frac{x^m}{m!}\hspace{1cm}\text{where~} |x|<1
	\end{align}
	
	There is only one lower Pochhammer parameter and it is not singular. Hence this Gauss function has Taylor expansion around $\e=0$. So we do not need any step down operators in this example. The coefficients of $\e^i$ are found by taking successive derivative with respect to $\e$ (\eeqref{eqn:TaylorExp}).  Using \eeqref{eqn:deriformula} we find the expansion 
	\begin{align}\label{eqn:2f1halfinteger}
		&G = \, _2F_1\left(2,3;-\frac{3}{2};x\right)\nonumber\\
		&  -\frac{8 x}{3} \e~F\left[ \begin{array}{c}\{4,\{1,1\}\}, \{3,\{1,1\}\}, \{1,\{1,0\}\}, \{1,\{0,1\}\}, \left\{-\frac{3}{2},\{0,1\}\right\}\\
			\{2,\{1,1\}\}, \left\{-\frac{1}{2},\{1,1\}\right\}, \left\{-\frac{1}{2},\{0,1\}\right\}\end{array}\;\middle|\;
		\{x,x\} \right] \nonumber\\
		&-\frac{16 x}{9} \e^2~F\left[ \begin{array}{c}\{4,\{1,1\}\}, \{3,\{1,1\}\}, \{1,\{1,0\}\}, \{1,\{0,1\}\}, \left\{-\frac{3}{2},\{0,1\}\right\}\\
			\{2,\{1,1\}\}, \left\{-\frac{1}{2},\{1,1\}\right\}, \left\{-\frac{1}{2},\{0,1\}\right\}\end{array}\;\middle|\;
		\{x,x\} \right] +192 x^2 \e^2\nonumber\\
		&\times F\left[ \begin{array}{c}\{5,\{1,1,1\}\}, \{4,\{1,1,1\}\}, \{1,\{1,0,0\}\}, \left\{-\frac{1}{2},\{0,1,1\}\right\}, \{1,\{0,1,0\}\}, \{1,\{0,0,1\}\}, \left\{-\frac{1}{2},\{0,0,1\}\right\}\\
			\{3,\{1,1,1\}\}, \left\{\frac{1}{2},\{1,1,1\}\right\}, \left\{\frac{1}{2},\{0,1,1\}\right\}, \left\{\frac{1}{2},\{0,0,1\}\right\}\end{array}\;\middle|\;
		\{x,x,x\} \right]\nonumber\\
		&+ O(\e^3)
	\end{align}
	The coefficients of $\e^3$ and $\e^4$ can be found in the ancillary file. We show the Taylor expansion obtained using \texttt{HypExp} \cite{HypExp,HypExp2} below for comparison.
	\begin{align}\label{eqn:2f1halfintegerHypExp}
		G&= \frac{-315 i \sqrt{-\frac{x}{x-1}} (4 x+7) x^2 ~\text{HPL}\left(\{\text{plus}\},i \sqrt{-\frac{x}{x-1}}\right)+140 x^4+4294 x^3+2784 x^2-320 x+32}{32 (x-1)^6}\nonumber\\
		&+\frac{\e x}{48 (x-1)^6} \Bigg[  12 i (143 x+329) \sqrt{-\frac{x}{x-1}} x \text{HPL}\left(\{\text{plus}\},i \sqrt{-\frac{x}{x-1}}\right)\nonumber\\
		&+945 i (4 x+7) \sqrt{-\frac{x}{x-1}} x \text{HPL}\left(\{0,\text{plus}\},i \sqrt{-\frac{x}{x-1}}\right)-144 x^3+2822 x^2+6912 x-128 \Bigg]\nonumber\\
		&+\frac{\e^2 x}{72 (x-1)^6}\Bigg[
		-11340 i \sqrt{-\frac{x}{x-1}} x^2 \text{HPL}\left(\{0,0,\text{plus}\},i \sqrt{-\frac{x}{x-1}}\right)\nonumber\\
		&-18 i \sqrt{-\frac{x}{x-1}} (42 x+145) x \text{HPL}\left(\{\text{plus}\},i \sqrt{-\frac{x}{x-1}}\right)-36 i \sqrt{-\frac{x}{x-1}} (143 x+329) x \text{HPL}\left(\{0,\text{plus}\},i \sqrt{-\frac{x}{x-1}}\right)\nonumber\\
		&-19845 i \sqrt{-\frac{x}{x-1}} x \text{HPL}\left(\{0,0,\text{plus}\},i \sqrt{-\frac{x}{x-1}}\right)+36 x^3+14474 x^2+20736 x-128
		\Bigg]
	\end{align}
Here HPL's are harmonic polylogarithms \cite{Remiddi:1999ew,Maitre2006}. As our motivation is to keep the hypergeometric structure intact throughout the calculation, we do not find equivalence of   \eeqref{eqn:2f1halfinteger} and \eeqref{eqn:2f1halfintegerHypExp} analytically. Although, we find excellent numerical agreement of the two results.  The numerical comparison of \eeqref{eqn:2f1halfinteger} and \eeqref{eqn:2f1halfintegerHypExp} are shown in Appendix  \ref{sec:Gausshalfintegernum} at a certain $x (<1)$ and found to be consistent.

	\subsection{Gauss $_2F_1$ example}\label{sec:Gauss}
	Let us consider the another example of \Ghf
	\begin{align} \label{eqn:Gaussinteger}
		G \vcentcolon =~ _2F_1(\e,-\e;\e-1;x)= \sum_{m=0}^\infty \frac{ (\e)_m (-\e)_m}{ (\e-1)_m} \frac{x^m}{m!} \hspace{1cm}\text{where~} |x|<1
	\end{align}
Here the lower Pochhammer parameter is singular so we find the step down operator first. This will allow us to relate the above function with another \hf which can be expanded in Taylor expansion.
	\begin{align} \label{eqn:Gaussstepdown}
		G  = H \bullet~ _2F_1(\e,-\e;\e+1;x)
	\end{align}
We find the step down operator $H$ to be,
\begin{align}\label{eqn:Gaussstepdownop}
	H &= \frac{(x (\e (2 x-1)-x+1))}{(\e-1) \e (x-1)} \partial_x +\frac{\e (2 x-1)-x+1}{(\e-1) (x-1)}\nonumber\\
	&= \frac{1}{\e} x \partial_x + \e^0 \left(1-\frac{x^2 \partial_x}{x-1}\right) + \e \left(-\frac{x \left(x \partial_x+1\right)}{x-1}\right) + \e^2 \left(-\frac{ x \left(x \partial_x+1\right)}{x-1} \right) + \e^3 \left(-\frac{ x \left(x \partial_x+1\right)}{x-1}\right) + O(\e^4)
\end{align}	
In the last step, we expanded the step down operator in series of $\e$.

Next we find the Taylor expansion of $_2F_1$ appearing in the right hand side of \eeqref{eqn:Gaussstepdown}.
\begin{align} \label{eqn:GaussTaylor}
_2F_1(\e,-\e;\e+1;x) &= 1 - \e^2 \left(x~ \, _3F_2(1,1,1;2,2;x)\right) 
+ \e^3 \Bigg(\frac{2}{3} x \, _3F_2(1,1,1;2,2;x) \nonumber\\
&+\frac{x}{3}~F\left[ \begin{array}{c}\{1,\{1,1\}\}, \{1,\{1,1\}\}, \{1,\{1,0\}\}, \{1,\{0,1\}\}, \{1,\{0,1\}\}\\
	\{2,\{1,1\}\}, \{2,\{1,1\}\}, \{2,\{0,1\}\}\end{array}\;\middle|\;
\{x,x\} \right]\nonumber\\
&+ \frac{x^2}{12}~F\left[ \begin{array}{c}\{2,\{1,1\}\}, \{2,\{1,1\}\}, \{1,\{1,0\}\}, \{2,\{0,1\}\}, \{1,\{0,1\}\}\\
	\{3,\{1,1\}\}, \{3,\{1,1\}\}, \{3,\{0,1\}\}\end{array}\;\middle|\;
\{x,x\} \right]
\Bigg) + O(\e^4)
\end{align}
The coefficient of $\e^4$  can be found in the ancillary file.

Now applying the step down operator (i.e. \eeqref{eqn:Gaussstepdownop}) on the Taylor series (\eeqref{eqn:GaussTaylor}), we find the series expansion of the Gauss $_2F_1$ we started with (i.e. \eeqref{eqn:Gaussinteger}),
\begin{align} \label{eqn:Gaussresult}
G &= 1 + \e \left(-\frac{x}{x-1} -x \, _3F_2(1,1,1;2,2;x) -  \frac{x^2}{4} \, _3F_2(2,2,2;3,3;x)\right)\nonumber\\
&+ \e^2 \Bigg[ -\frac{x}{x-1} + \frac{(x (2 x+1)) }{3 (x-1)} \, _3F_2(1,1,1;2,2;x) + \frac{\left(x^2 (5 x-2)\right) }{12 (x-1)} \, _3F_2(2,2,2;3,3;x)\nonumber\\
&+ \frac{x}{3}~F\left[ \begin{array}{c}\{1,\{1,1\}\}, \{1,\{1,1\}\}, \{1,\{1,0\}\}, \{1,\{0,1\}\}, \{1,\{0,1\}\}\\
	\{2,\{1,1\}\}, \{2,\{1,1\}\}, \{2,\{0,1\}\}\end{array}\;\middle|\;
\{x,x\} \right]\nonumber\\
&+ \frac{x^2}{24}~F\left[ \begin{array}{c}\{2,\{1,1\}\}, \{2,\{1,1\}\}, \{1,\{1,0\}\}, \{2,\{0,1\}\}, \{2,\{0,1\}\}\\
	\{3,\{1,1\}\}, \{3,\{1,1\}\}, \{3,\{0,1\}\}\end{array}\;\middle|\;
\{x,x\} \right]\nonumber\\
&+ \frac{x^2}{12}~F\left[ \begin{array}{c}\{2,\{1,1\}\}, \{2,\{1,1\}\}, \{2,\{1,0\}\}, \{1,\{0,1\}\}, \{1,\{0,1\}\}\\
	\{3,\{1,1\}\}, \{3,\{1,1\}\}, \{2,\{0,1\}\}\end{array}\;\middle|\;
\{x,x\} \right]\nonumber\\
&+ \frac{x^2}{6}~F\left[ \begin{array}{c}\{2,\{1,1\}\}, \{2,\{1,1\}\}, \{1,\{1,0\}\}, \{2,\{0,1\}\}, \{1,\{0,1\}\}\\
	\{3,\{1,1\}\}, \{3,\{1,1\}\}, \{3,\{0,1\}\}\end{array}\;\middle|\;
\{x,x\} \right]\nonumber\\
&+ \frac{2 x^3}{81}~F\left[ \begin{array}{c}\{3,\{1,1\}\}, \{3,\{1,1\}\}, \{1,\{1,0\}\}, \{3,\{0,1\}\}, \{2,\{0,1\}\}\\
	\{4,\{1,1\}\}, \{4,\{1,1\}\}, \{4,\{0,1\}\}\end{array}\;\middle|\;
\{x,x\} \right]\nonumber\\
&+ \frac{x^3}{27}~F\left[ \begin{array}{c}\{3,\{1,1\}\}, \{3,\{1,1\}\}, \{2,\{1,0\}\}, \{2,\{0,1\}\}, \{1,\{0,1\}\}\\
	\{4,\{1,1\}\}, \{4,\{1,1\}\}, \{3,\{0,1\}\}\end{array}\;\middle|\;
\{x,x\} \right]  \Bigg] + O(\e^3)
\end{align}
We present the series expansion obtained from \texttt{HypExp} package \cite{HypExp} \cite{HypExp2} for comparison,
\begin{align} \label{eqn:GaussresultHypExp}
	G = 1 + \e \left(\log (1-x)-\frac{x}{x-1} \right) + \e^2 \left(-\frac{x}{x-1}+\frac{1}{2} \log ^2(1-x)-\frac{x \log (1-x)}{x-1}\right) + O(\e^3)
\end{align}
Our result (\eeqref{eqn:Gaussresult}) is found to be in agreement with the result from \texttt{HypExp} (\eeqref{eqn:GaussresultHypExp}) numerically at given $x$ ($<1$) evaluated with sufficient number of terms. It is also consistent with \texttt{NumExp} package \cite{NumExp}. The numerical comparison can be found in Appendix \ref{sec:Gaussinteger}.

\subsection{Taylor series of Appell $F_1$: An application to one loop massive vertex integral}

In the paper \cite{Phan:2018cnz}, one loop one- to four-point functions are obtained as a meromorphic function of the space-time dimension $d$. In particular, the one loop massive vertex integral is expressed in terms of double variable Appell $F_1$ function (Eq. B.2 of the mentioned paper) 

\begin{align}
	F_1 \left(\frac{d-2}{2},1,\frac{1}{2}, \frac{d}{2}, x_c,y_c\right)
\end{align}
 The authors also provided a table (Table B.1) for numerical evaluation of the Appell function $F_1$. The numerical evaluation of Appell $F_1$ function for complex values of its argument is the subject of our forthcoming article \cite{ASS}. In this section, we focus on the $\e$-expansion of the above Appell  function. With $d=4-2\e$, the function becomes
 \begin{align}\label{eqn:PhanF1}
		F_1 \left(1-\e,1,\frac{1}{2},2-\e, x,y\right)
 \end{align}
The numerical values of the coefficients of the series expansion of the $F_1$ (\eeqref{eqn:PhanF1}) is given in Table B.2 of \cite{Phan:2018cnz} at $(x_c,y_c) = (11.1-10^{-12} i, 12.1 -10^{-12} i)$. Since this point does not lie in the defining domain of convergence of the Appell $F_1$ function, which is $|x|<1 \wedge |y|<1$, we find a proper analytic continuation before carrying out the series expansion. One such analytic continuation of $F_1$ from \cite{ASS} is written below, skipping the derivation, which is valid at the aforementioned values of $(x_c, y_c)$.
\begin{align}\label{eqn:AppellF1AC}
	F_1(a,b_1,b_2,c,x,y)&=(1-x)^{-a}\left(\frac{1}{1-x}\right)^{-a+b_1+b_2}\frac{ \Gamma (c)  \Gamma \left(a-b_1-b_2\right)}{\Gamma (a) \Gamma \left(c-b_1-b_2\right)}\nonumber \\ 
	&\times F_2\left(b_1+b_2,c-a,b_2,-a+b_1+b_2+1,b_1+b_2,\frac{1}{1-x},\frac{x-y}{x-1}\right)\nonumber\\
	&+ (1-x)^{-a} \frac{ \Gamma (c) \Gamma \left(-a+b_1+b_2\right)}{\Gamma \left(b_1+b_2\right) \Gamma (c-a)}\nonumber\\
	&\times F_2\left(a,-b_1-b_2+c,b_2,a-b_1-b_2+1,b_1+b_2,\frac{1}{1-x},\frac{x-y}{x-1}\right)
\end{align}
with the domain of convergence
\begin{align}
	| 1-x| >1\land \left| \frac{x-y}{x-1}\right| <1\land \left| \frac{x-y}{x-1}\right| +\frac{1}{| 1-x| }<1
\end{align}
where the functions appearing in the RHS of \eeqref{eqn:AppellF1AC} are Appell $F_2$  (see Appendix \ref{Appendix:notation}). Applying the analytic continuation on the $F_1$ function in \eeqref{eqn:PhanF1} we find
\begin{align}\label{eqn:Phanpreexpansion}
	F_1 &(1-\e,1,\frac{1}{2},2-\e, x,y) \nonumber\\ &=\frac{\left(\frac{1}{1-x}\right)^{\epsilon +\frac{1}{2}} (1-x)^{\epsilon -1} \Gamma \left(-\epsilon -\frac{1}{2}\right) \Gamma (2-\epsilon )}{\Gamma \left(\frac{1}{2}-\epsilon \right) \Gamma (1-\epsilon )} 
	\sum_{m,n=0}^\infty \frac{(1)_m  \left(\frac{1}{2}\right)_n \left(\frac{3}{2}\right)_{m+n} }{m! n! \left(\frac{3}{2}\right)_n \left(\epsilon +\frac{3}{2}\right)_m} \left(\frac{1}{1-x}\right)^m \left(\frac{x-y}{x-1}\right)^n\nonumber\\
	&+ \frac{2 (1-x)^{\epsilon -1} \Gamma (2-\epsilon ) \Gamma \left(\epsilon +\frac{1}{2}\right)}{\sqrt{\pi }} \sum_{m,n=0}^\infty \frac{ \left(\frac{1}{2}\right)_n  (1-\epsilon )_{m+n}}{m! n! \left(\frac{3}{2}\right)_n} \left(\frac{1}{1-x}\right)^m \left(\frac{x-y}{x-1}\right)^n \\
	& = g_1 (x,y,\e) \times S_1 + g_2 (x,y,\e) \times S_2
\end{align}
Now we are ready to perform the series expansion of the above expression. The prefactors $g_1$ and $g_2$ can be easily expanded in series of $\e$, where as the series $S_1$ and $S_2$ can be expanded in Taylor series in $\e$ following the prescription of Section \ref{sec:TaylorSeries}. The first few terms in the series expansion of the prefactors $g_1, g_2$ read,
\begin{align}
	&g_1 = -2 \left(\frac{1}{1-x}\right)^{3/2}+ \e \left[\frac{2 \sqrt{\frac{1}{1-x}} \left(\log \left(\frac{1}{1-x}\right)+\log (1-x)-3\right)}{x-1}\right] + O(\e^2)\\
	&g_2 = -\frac{2}{x-1} +  \e \big[\frac{-2 \log (1-x)+2+\log (16)}{x-1}\big] + O(\e^2)
\end{align}
and the first few terms of the expansion of the series read
\begin{align}
	S_1 &= \sum_{m,n=0}^\infty \frac{(1)_m  \left(\frac{1}{2}\right)_n \left(\frac{3}{2}\right)_{m+n} }{m! n! \left(\frac{3}{2}\right)_m \left(\frac{3}{2}\right)_n} \left(\frac{1}{1-x}\right)^m \left(\frac{x-y}{x-1}\right)^n\nonumber\\
	&+  \e \left[ \frac{2 }{3 (x-1)} \sum_{m,n,p=0}^\infty \frac{(1)_m \left(\frac{1}{2}\right)_n (1)_p \left(\frac{3}{2}\right)_p  \left(\frac{5}{2}\right)_{m+n+p}}{ m! n! p! \left(\frac{3}{2}\right)_n \left(\frac{5}{2}\right)_p \left(\frac{5}{2}\right)_{m+p}} \left(\frac{1}{1-x}\right)^{m+p} \left(\frac{x-y}{x-1}\right)^n \right] +O(\e^2)\nonumber\\
	S_2 &= \sum_{m,n=0} \frac{ \left(\frac{1}{2}\right)_n (1)_{m+n} }{m! n! \left(\frac{3}{2}\right)_n} \left(\frac{1}{1-x}\right)^m \left(\frac{x-y}{x-1}\right)^n \\
	&+ \e \Bigg[ \frac{(x-y)}{(3-3 x)} \sum_{m,n,p=0}^\infty\frac{ (1)_n \left((1)_p\right){}^2  \left(\frac{3}{2}\right)_{n+p} (2)_{m+n+p} }{ m! n! p! (2)_p (2)_{n+p} \left(\frac{5}{2}\right)_{n+p}} \left(\frac{1}{1-x}\right)^m \left(\frac{x-y}{x-1}\right)^{n+p}\nonumber\\
	& -\frac{1}{(1-x) }\sum_{m,n,p=0}^\infty \frac{(1)_m \left(\frac{1}{2}\right)_n (1)_p  (1)_{n+p}  (2)_{m+n+p}}{m! n! p! \left(\frac{3}{2}\right)_n (2)_{m+p} (2)_{n+p}} \left(\frac{1}{1-x}\right)^{m+p} \left(\frac{x-y}{x-1}\right)^n \Bigg] + O(\e^2)
\end{align}
 Combining the expansion of the prefactors and the series, we write the expansion of the Appell $F_1$ function \eeqref{eqn:PhanF1} in the notation described in Appendix \eqref{Appendix:notation}
\begin{align}\label{eqn:PhanAppellF1exp}
F_1 &(1-\e,1,\frac{1}{2},2-\e, x,y) = \nonumber\\
& -2 \left(\frac{1}{1-x}\right)^{3/2}~F\left[ \
\begin{array}{c}\left\{\frac{3}{2},\{1,1\}\right\}, \{1,\{1,0\}\}, \
	\left\{\frac{1}{2},\{0,1\}\right\}\\
	\left\{\frac{3}{2},\{1,0\}\right\}, \left\{\frac{3}{2},\{0,1\}\right\}\end{array}\;\middle|\;
\left\{\frac{1}{1-x},\frac{x-y}{x-1}\right\} \right]\nonumber\\
& -\frac{2}{x-1}~F\left[ \begin{array}{c}\{1,\{1,1\}\}, \left\{\frac{1}{2},\{0,1\}\right\}\\
	\left\{\frac{3}{2},\{0,1\}\right\}\end{array}\;\middle|\;
\left\{\frac{1}{1-x},\frac{x-y}{x-1}\right\} \right] \nonumber\\
&+ \e \Bigg[  \frac{2 \sqrt{\frac{1}{1-x}} \left(\log \left(\frac{1}{1-x}\right)+\log (1-x)-3\right)}{x-1}~F\left[ \
\begin{array}{c}\left\{\frac{3}{2},\{1,1\}\right\}, \{1,\{1,0\}\}, \
	\left\{\frac{1}{2},\{0,1\}\right\}\\
	\left\{\frac{3}{2},\{1,0\}\right\}, \left\{\frac{3}{2},\{0,1\}\right\}\end{array}\;\middle|\;
\left\{\frac{1}{1-x},\frac{x-y}{x-1}\right\} \right]\nonumber\\
&-\frac{4 \left(\frac{1}{1-x}\right)^{3/2}}{3 (x-1)}~F\left[ \begin{array}{c}\left\{\frac{5}{2},\{1,1,1\}\right\}, \{1,\{1,0,0\}\}, \left\{\frac{1}{2},\{0,1,0\}\right\}, \left\{\frac{3}{2},\{0,0,1\}\right\}, \{1,\{0,0,1\}\}\\
	\left\{\frac{5}{2},\{1,0,1\}\right\}, \left\{\frac{3}{2},\{0,1,0\}\right\}, \left\{\frac{5}{2},\{0,0,1\}\right\}\end{array}\;\middle|\;
\left\{\frac{1}{1-x},\frac{x-y}{x-1},\frac{1}{1-x}\right\} \right]\nonumber\\
&-\frac{2}{(x-1)^2}~F\left[ \begin{array}{c}\{2,\{1,1,1\}\}, \
	\{1,\{1,0,0\}\}, \{1,\{0,1,1\}\}, \left\{\frac{1}{2},\{0,1,0\}\right\}, \{1,\{0,0,1\}\}\\
	\{2,\{1,0,1\}\}, \{2,\{0,1,1\}\}, \left\{\frac{3}{2},\{0,1,0\}\right\}\end{array}\;\middle|\;
\left\{\frac{1}{1-x},\frac{x-y}{x-1},\frac{1}{1-x}\right\} \right] \nonumber\\
& +\frac{2 (x-y)}{3 (x-1)^2}~F\left[ \
\begin{array}{c}\{2,\{1,1,1\}\}, \left\{\frac{3}{2},\{0,1,1\}\right\}, \{1,\{0,1,0\}\}, \{1,\{0,0,1\}\}, \{1,\{0,0,1\}\}\\
	\left\{\frac{5}{2},\{0,1,1\}\right\}, \{2,\{0,1,1\}\}, \
	\{2,\{0,0,1\}\}\end{array}\;\middle|\;
\left\{\frac{1}{1-x},\frac{x-y}{x-1},\frac{x-y}{x-1}\right\} \right]\nonumber\\
&-\frac{2 \left(\log \left(\frac{1-x}{4}\right)-1\right)}{x-1}~F\left[ \
\begin{array}{c}\{1,\{1,1\}\}, \left\{\frac{1}{2},\{0,1\}\right\}\\
	\left\{\frac{3}{2},\{0,1\}\right\}\end{array}\;\middle|\;
\left\{\frac{1}{1-x},\frac{x-y}{x-1}\right\} \right]\Bigg] + O(\e^2)
\end{align}
The higher order terms of the expansion can be found in the ancillary notebook. A numerical comparison of the expansion coefficients of $F_1$ (\eeqref{eqn:PhanAppellF1exp}) found above with the results given in Table B.2 of \cite{Phan:2018cnz} can be found in Appendix \ref{sec:PhanF1}.

	\subsection{Laurent series of Appell $F_1$ }\label{sec:AppellF1example}
	Let us consider the following double variable Appell $F_1$ function which we wish to find the $\e$ expansion of, 
	\begin{align}\label{eqn:AppellF1v1}
		F_1(3/2 , 2\e +1, 4-\e, \e-2,x,y) = \sum_{m,n=0}^\infty \frac{ \left(\frac{3}{2}\right)_{m+n} (2 \e+1)_m (4-\e)_n }{(\e-2)_{m+n}} \frac{x^m }{m! } \frac{y^n}{ n!} \hspace{1cm}\text{where~~} |x|<1 \wedge |y|<1
	\end{align}
	The corresponding Appell $F_1$ function that can be expanded in Taylor series is given by
	\begin{align}\label{eqn:AppellF1v2}
		F_1(3/2 , 2\e +1, 4-\e, \e+1,x,y) = \sum_{m,n=0}^\infty \frac{ \left(\frac{3}{2}\right)_{m+n} (2 \e+1)_m (4-\e)_n }{(\e+1)_{m+n}} \frac{x^m }{m! } \frac{y^n}{ n!} \hspace{1cm}\text{where~~} |x|<1 \wedge |y|<1
	\end{align}
We find the Taylor expansion of the Appell $F_1$ function in \eeqref{eqn:AppellF1v2} and act the step down operator on it to find Laurent series expansion of the Appell $F_1$ function \eeqref{eqn:AppellF1v1}.

The step down operator that relates the above Appell $F_1$ function in \eeqref{eqn:AppellF1v1} with the Appell $F_1$ function in \eeqref{eqn:AppellF1v2} is found to be,
\begin{align}
	H = C_0 + C_1 \partial_x + C_2 \partial_y
\end{align}
where are rational functions $C_i$'s are 

\begin{align}
	C_0 &=\frac{1}{4 \e \left(\e^2-3 \e+2\right) (x-1)^2 (y-1)^2}\Big[4 \e^3 (x-1)^2 (y-1)^2-12 \e^2 \left(x^2 \left(2 y^2-5 y+2\right)-2 x \left(y^2-4 y+2\right)-y^2-y+1\right)\nonumber\\
	&+\e \left(x^2 \left(5 y^2-160 y+68\right)+2 x \left(91 y^2+10 y-14\right)-139 y^2+44 y+8\right)\nonumber\\
	&+3 \left(x^2 \left(95 y^2-54 y+11\right)-4 x \left(39 y^2-14 y+1\right)+4 y (17 y-4)\right) \Big]\\
	C_1 &=\frac{1}{4 (\e-2) (\e-1) \e (x-1)^2 (y-1)^2} \Big[x  \Big(4 \e^2 \left(x^2 \left(y^2+1\right)+x \left(-4 y^2+y-1\right)+4 y^2-3 y+1\right)\nonumber\\
	&-2 \e \left(x^2 \left(22 y^2+6 y+1\right)+x \left(-78 y^2+39 y-19\right)+44 y^2-21 y+6\right)+x^2 \left(225 y^2-178 y+57\right)\nonumber\\
	&-2 x \left(155 y^2-66 y+15\right)+8 \left(15 y^2-3 y+1\right)\Big) \Big]\\
	C_2 &= \frac{1}{4 (\e-2) (\e-1) \e (x-1)^2 (y-1)^2}\Big[y \Big(4 \e^2 \left(x^2 (y+1)^2-2 x (3 y+1) y+7 y^2-4 y+1\right)\nonumber\\
	&-4 \e \left(x^2 \left(11 y^2+18 y-5\right)-6 x \left(9 y^2-2 y+1\right)+35 y^2-14 y+3\right)\nonumber\\
	&+x^2 \left(225 y^2-106 y+24\right)-2 x \left(191 y^2-57 y+9\right)+171 y^2-36 y+8\Big) 
	\Big]
\end{align}
Now expanding all the three coefficients $C_i$, in $\e$, we club  coefficients of different powers of $\e$ together,
\begin{align} \label{eqn:HAppellF1}
	H =  A_1 \frac{1}{\e} + A_2 \e^0 + A_3 \e+ A_4 \e^2 + O(\e^3)
\end{align}
where,
\begin{align}
	A_1  &=\frac{x  \left(x^2 \left(225 y^2-178 y+57\right)-2 x \left(155 y^2-66 y+15\right)+8 \left(15 y^2-3 y+1\right)\right) \partial_x}{8  (x-1)^2 (y-1)^2 }\nonumber\\
	&+\frac{y \left(x^2 \left(225 y^2-106 y+24\right)-2 x \left(191 y^2-57 y+9\right)+171 y^2-36 y+8\right)  \partial_y}{8  (x-1)^2 (y-1)^2}\nonumber\\
	&+\frac{3 \left(x^2 \left(95 y^2-54 y+11\right)-4 x \left(39 y^2-14 y+1\right)+4 y (17 y-4)\right)}{8  (x-1)^2 (y-1)^2}   \label{eqn:AppellF1A1}\\
	A_2 &= \frac{x  \left(587 x^2 y^2-558 x^2 y+167 x^2-618 x y^2+240 x y-14 x+184 y^2+12 y\right) \partial_x}{16 (x-1)^2 (y-1)^2}\nonumber\\
	&+\frac{y \left(587 x^2 y^2-462 x^2 y+112 x^2-714 x y^2+246 x y-6 x+233 y^2+4 y\right) \partial_y}{16 (x-1)^2 (y-1)^2}\nonumber\\
	&+ \frac{865 x^2 y^2-806 x^2 y+235 x^2-1040 x y^2+544 x y-92 x+334 y^2-56 y+16}{16 (x-1)^2 (y-1)^2}  \label{eqn:AppellF1A2}\\
	A_3  &=\frac{x  \left(1327 x^2 y^2-1318 x^2 y+403 x^2-1298 x y^2+472 x y+2 x+376 y^2+36 y\right) \partial_x}{32 (x-1)^2 (y-1)^2}\nonumber\\
	&+\frac{y  \left(1327 x^2 y^2-1142 x^2 y+304 x^2-1474 x y^2+478 x y+18 x+469 y^2+20 y\right) \partial_y}{32 (x-1)^2 (y-1)^2}\nonumber\\
	&+\frac{3 \left(643 x^2 y^2-618 x^2 y+181 x^2-696 x y^2+304 x y-20 x+214 y^2-8 y\right)}{32 (x-1)^2 (y-1)^2}
\end{align}
The Taylor Expansion of the Appell $F_1$ in \eeqref{eqn:AppellF1v2} is found taking successive derivative with respect to $\e$ i.e.,
\begin{align}\label{eqn:TaylorF1}
F_1(\e,x,y) &= F_1(0,x,y) + \e F_1'(\e,x,y)\Big\vert_{\e=0} + \frac{1}{2!}\e^2 F_1''(\e,x,y)\Big\vert_{\e=0} + \frac{1}{3!} \e^3 F_1'''(\e,x,y)\Big\vert_{\e=0}+ O(\e^4)\\
&= f_0 + \e f_1 +\e^2 f_2 +  \e^3 f_3 +  O(\e^4) \label{eqn:TaylorF1v2}
\end{align}
All the Pochhammer parameters of the Appell $F_1$ in \eeqref{eqn:AppellF1v2} are collectively denoted as $\e$ in the argument of $F_1$ in \eeqref{eqn:TaylorF1} to give emphasis on the fact that the parameters are $\e$ dependent. The primes in the right hand side refer to the derivative with respect to $\e$. We write the first few $f_i$ below.
\begin{align}
	f_0 &= F_1\left(\frac{3}{2},1,4,1,x,y\right) \label{eqn:ApellF1f0}\\
	f_1 &=3 x~F\left[ \begin{array}{c}\left\{\frac{5}{2},\{1,1,1\}\right\}, \{1,\{1,0,0\}\}, \{1,\{1,0,0\}\}, \{1,\{0,1,0\}\}, \{4,\{0,0,1\}\}\\
		\{2,\{1,1,1\}\}, \{2,\{1,0,0\}\}\end{array}\;\middle|\;
	\{x,x,y\} \right]\nonumber\\
	&-\frac{3 x}{2}~F\left[ \begin{array}{c}\left\{\frac{5}{2},\{1,1,1\}\right\}, \{1,\{1,0,1\}\}, \{1,\{1,0,0\}\}, \{1,\{0,1,0\}\}, \{4,\{0,0,1\}\}\\
		\{2,\{1,1,1\}\}, \{2,\{1,0,1\}\}\end{array}\;\middle|\;
	\{x,x,y\} \right]\nonumber\\
	&-6 y~F\left[ \begin{array}{c}\left\{\frac{5}{2},\{1,1,1\}\right\}, \{5,\{1,0,1\}\}, \{1,\{1,0,0\}\}, \{1,\{1,0,0\}\}, \{1,\{0,1,0\}\}, \{1,\{0,0,1\}\}\\
		\{2,\{1,1,1\}\}, \{2,\{1,0,1\}\}, \{2,\{1,0,0\}\}\end{array}\;\middle|\;
	\{y,x,y\} \right]\nonumber\\
	&-\frac{3 y}{2}~F\left[ \begin{array}{c}\left\{\frac{5}{2},\{1,1,1\}\right\}, \{5,\{1,0,1\}\}, \{4,\{1,0,0\}\}, \{1,\{1,0,0\}\}, \{1,\{0,1,0\}\}, \{1,\{0,0,1\}\}\\
		\{2,\{1,1,1\}\}, \{2,\{1,0,1\}\}, \{5,\{1,0,0\}\}\end{array}\;\middle|\;
	\{y,x,y\} \right] \label{eqn:ApellF1f1}
\end{align}
We do not show the functions $f_2, f_3$ here due to its long expression. It is provided in the ancillary notebook.

Now applying the step down operator on the Appell $F_1$ in \eeqref{eqn:AppellF1v2},
\begin{align}
	F_1(3/2 , 2\e +1, 4-\e, \e-2,x,y) = H \bullet F_1(3/2 , 2\e +1, 4-\e, \e+1,x,y)
\end{align}
Using \eeqref{eqn:HAppellF1} and \eeqref{eqn:TaylorF1v2} on the right hand side of the above equation, 
\begin{align}\label{eqn:ApellF1Laurent}
F_1(3/2 , 2\e +1, &4-\e, \e-2,x,y)\nonumber\\
&= \left( A_1 \frac{1}{\e} + A_2 \e^0 + A_3 \e+ A_4 \e^2 + O(\e^3) \right) \bullet \left( f_0 + \e f_1 + \e^2 f_2 + \e^3 f_3 + O(\e^4)\right)\nonumber\\
&= \frac{1}{\e} \left[A_1 \bullet f_0\right] + \left[A_1 \bullet f_1 + A_2 \bullet f_0 \right]+ \e \left[A_3 \bullet f_0 + A_2 \bullet f_1 + A_1 \bullet f_2\right] \nonumber\\
 &+ \e^2 \left[A_4 \bullet f_0 + A_3 \bullet f_1 + A_2 \bullet f_2 + A_1 \bullet f_3\right] + O(\e^4)
\end{align}
We apply $A_1$ (\eeqref{eqn:AppellF1A1}) on $f_0$ (\eeqref{eqn:ApellF1f0}) to find the coefficient of $\e^{-1}$ of the Laurent expansion,
\begin{align}
A_1 \bullet f_0 &= \frac{3 \left(x^2 \left(95 y^2-54 y+11\right)-4 x \left(39 y^2-14 y+1\right)+4 y (17 y-4)\right)}{8  (x-1)^2 (y-1)^2} ~ F_1\left(\frac{3}{2},1,4,1,x,y\right)\nonumber\\
&+\frac{3 x \left(x^2 \left(225 y^2-178 y+57\right)-2 x \left(155 y^2-66 y+15\right)+8 \left(15 y^2-3 y+1\right)\right)}{16  (x-1)^2 (y-1)^2}   ~ F_1\left(\frac{5}{2},2,4,2,x,y\right)\nonumber\\
&+\frac{3 y \left(x^2 \left(225 y^2-106 y+24\right)-2 x \left(191 y^2-57 y+9\right)+171 y^2-36 y+8\right)}{4  (x-1)^2 (y-1)^2} ~ F_1\left(\frac{5}{2},1,5,2,x,y\right)
\end{align}
Similarly one find the higher order term of the Laurent series. The coefficient of $\e^0$,$\e^1$ in the Laurent series can be found in the ancillary notebook. The coefficients are numerically tested and found to be consistent. The results of the numerical tests can be found in the Appendix \ref{sec:AppellF1num}.

 \subsection{Appell $F_2$ example} \label{sec:MochF2}
 It is known that the result of the $\e$-expansion of Appell $F_2$ function around integer Pochhammer parameters can be expressed in terms of HPLs and  S-sums \cite{Moch:2001zr}. Algorithms are also provided to carry out such expansion in the mentioned article. The example given in Eq. (81) of \cite{Moch:2001zr} deals with the following Appell $F_2$ function
 \begin{align}\label{eqn:MochF2}
 	F_2(1,1,\e,1+\e,1-\e,x,y)
 \end{align}
However, we consider the Appell $F_2$ function with parameters $a_i, i =1,\dots,4$ in this example
\begin{align}\label{eqn:genF2}
	F_2 := F_2 (a_1, a_2 ,\e , a_3+\e, a_4-\e,x,y)
\end{align}
where it is assumed that $a_1,a_2 \in \mathbb{C}$ and $a_3,a_4 \in \mathbb{C}-\mathbb{Z}_{\le 0}$ in order to ensure that the function can be expanded in Taylor series. 
In this section, we find the series expansion of the above $F_2$ function \eeqref{eqn:genF2} and express the result in terms of MHFs. 

Following the prescription discussed in Section \ref{sec:TaylorSeries}, we write the first few terms of the series expansion below.

\begin{align}\label{eqn:genAppellF2exp}
	&F_2 = ~_2F_1 (a_1,a_2,a_3,x) +\nonumber\\
	&\e \Bigg[ -\frac{a_1 a_2 x}{a_3^2} \sum_{s_1,s_2 = 0}^\infty \frac{(1)_{s_1} (1)_{s_2}  \left(a_1+1\right)_{s_1+s_2} \left(a_2+1\right)_{s_1+s_2} \left(a_3\right)_{s_2}}{ (2)_{s_1+s_2} \left(a_3+1\right)_{s_2} \left(a_3+1\right)_{s_1+s_2}} \frac{x^{s_1+s_2}}{s_1! s_2!} \nonumber\\
	&+ \frac{a_1 y}{a_4} \sum_{s_1,s_2 = 0}^\infty \frac{\left((1)_{s_2}\right){}^2  \left(a_1+1\right)_{s_1+s_2} \left(a_2\right)_{s_1}}{ (2)_{s_2} \left(a_3\right)_{s_1} \left(a_4+1\right)_{s_2}} \frac{x^{s_1} y^{s_2}}{s_1! s_2!}
	\Bigg]\nonumber\\
	&+\e^2 \Bigg[ \frac{a_1 a_2 x}{a_3^3} \sum_{s_1,s_2 = 0}^\infty \frac{(1)_{s_1} (1)_{s_2}  \left(a_1+1\right)_{s_1+s_2} \left(a_2+1\right)_{s_1+s_2} \left(a_3\right)_{s_2}}{ (2)_{s_1+s_2} \left(a_3+1\right)_{s_2} \left(a_3+1\right)_{s_1+s_2}} \frac{x^{s_1+s_2}}{s_1! s_2!}\nonumber\\
	&+\frac{a_1 \left(a_1+1\right) a_2 \left(a_2+1\right) x^2}{2 a_3 \left(a_3+1\right){}^3}  \sum_{s_1,s_2,s_3 = 0}^\infty \frac{(1)_{s_1} (1)_{s_2} (1)_{s_3}  \left(a_1+2\right)_{s_1+s_2+s_3} \left(a_2+2\right)_{s_1+s_2+s_3} \left(a_3+1\right)_{s_3} \left(a_3+1\right)_{s_2+s_3}}{ (3)_{s_1+s_2+s_3} \left(a_3+2\right)_{s_3} \left(a_3+2\right)_{s_2+s_3} \left(a_3+2\right)_{s_1+s_2+s_3}} \frac{x^{s_1+s_2+s_3}}{s_1! s_2! s_3!}\nonumber\\
	&+\frac{a_1 y}{2 a_4^2}  \sum_{s_1,s_2 = 0}^\infty \frac{\left((1)_{s_2}\right){}^2  \left(a_1+1\right)_{s_1+s_2} \left(a_2\right)_{s_1}}{ (2)_{s_2} \left(a_3\right)_{s_1} \left(a_4+1\right)_{s_2}} \frac{x^{s_1} y^{s_2}}{s_1! s_2!} \nonumber\\
	&+\frac{a_1 y}{2 a_4^2} \sum_{s_1,s_2,s_3 = 0}^\infty \frac{(1)_{s_2} (1)_{s_3} (1)_{s_2+s_3}  \left(a_1+1\right)_{s_1+s_2+s_3} \left(a_2\right)_{s_1} \left(a_4\right)_{s_3}}{(2)_{s_2+s_3} \left(a_3\right)_{s_1} \left(a_4+1\right)_{s_3} \left(a_4+1\right)_{s_2+s_3}} \frac{x^{s_1} y^{s_2+s_3}}{s_1! s_2! s_3! }\nonumber\\
	&-\frac{a_1 \left(a_1+1\right) a_2 x y}{a_3^2 a_4} \sum_{s_1,s_2,s_3 = 0}^\infty \frac{(1)_{s_1} \left((1)_{s_2}\right){}^2 (1)_{s_3}  \left(a_1+2\right)_{s_1+s_2+s_3} \left(a_2+1\right)_{s_1+s_3} \left(a_3\right)_{s_3}}{ (2)_{s_2} (2)_{s_1+s_3} \left(a_3+1\right)_{s_3} \left(a_3+1\right)_{s_1+s_3} \left(a_4+1\right)_{s_2}} \frac{x^{s_1+s_3} y^{s_2}}{s_1! s_2! s_3!}\nonumber\\
	&+ \frac{a_1 \left(a_1+1\right) y^2}{4 a_4 \left(a_4+1\right){}^2}  \sum_{s_1,s_2,s_3 = 0}^\infty \frac{(1)_{s_2} (1)_{s_3} (2)_{s_2+s_3}  \left(a_1+2\right)_{s_1+s_2+s_3} \left(a_2\right)_{s_1} \left(a_4+1\right)_{s_3}}{ (3)_{s_2+s_3} \left(a_3\right)_{s_1} \left(a_4+2\right)_{s_3} \left(a_4+2\right)_{s_2+s_3}} \frac{x^{s_1} y^{s_2+s_3}}{s_1! s_2! s_3!}\nonumber\\
	&+\frac{a_1 y^2 \left(a_1 +1\right)}{a_4 \left(2 a_4+2\right)} \sum_{s_1,s_2,s_3 = 0}^\infty \frac{(1)_{s_2} \left((1)_{s_3}\right){}^2 (2)_{s_2+s_3}  \left(a_1+2\right)_{s_1+s_2+s_3} \left(a_2\right)_{s_1}}{ (2)_{s_3} (3)_{s_2+s_3} \left(a_3\right)_{s_1} \left(a_4+2\right)_{s_2+s_3}} \frac{x^{s_1} y^{s_2+s_3}}{s_1! s_2! s_3!}
	\Bigg] + O(\e^3)
\end{align}

The coefficient of $\e^3$ can be found in the ancillary notebook. The result obtained above is crosschecked numerically with Eq. (81) of \cite{Moch:2001zr} after putting $a_i$'s as $1$ in \eeqref{eqn:genAppellF2exp}. For the purpose of numerical evaluation \texttt{PolyLogTools} \cite{Duhr:2019tlz} is used. The results are found to be in good agreement. A table containing the numerical result is provided in Appendix \ref{sec:MochF2num}.

Furthermore, we take different sets of values (integers, half-integers, rationals) for the $a_i$'s in Appendix \ref{sec:parameterindependence} and show that, the convergences of the series coefficients are independent of the values of the Pochhammer parameters, as noted in \cite{Ancarani2017}.

\subsection{The 2-loop Sunset Integral in $D = 2$  dimension}
It is well known that, the 2-loop sunset integral with different internal masses is convergent in $D=2$ dimension. In this section we apply our methodology to find the analytic expression of the 2-loop sunset integral at $D=2$ dimension.
We use the expression in Eq. (16) of \cite{Berends:1993ee},
\begin{align}\label{eqn:Berendssunset}
	T_{123}(p^2,m_1^2,m_2^2,m_3^2) = -m_3^2 \left( \frac{m_3^2}{4 \pi \mu^2}\right)^{-2(\e+1)} \mathcal{S}
\end{align}
with,
\begin{align} \label{eqn:sunsetseries}
	\mathcal{S}&=z_1^{-\epsilon } z_2^{-\epsilon } \Gamma (\epsilon )^2 ~ F_C^{(3)}\left(1,1-\epsilon ,1-\epsilon ,1-\epsilon ,1-\epsilon ,z_1,z_2,z_3\right)\nonumber\\
	&-z_1^{-\epsilon } \Gamma (\epsilon )^2 ~F_C^{(3)}\left(1,\epsilon +1,1-\epsilon ,\epsilon +1,1-\epsilon ,z_1,z_2,z_3\right)\nonumber\\
	&-z_2^{-\epsilon } \Gamma (\epsilon )^2~ F_C^{(3)}\left(1,\epsilon +1,\epsilon +1,1-\epsilon ,1-\epsilon ,z_1,z_2,z_3\right)\nonumber\\
	&-\Gamma (-\epsilon ) \Gamma (\epsilon ) \Gamma (2 \epsilon +1)~ F_C^{(3)}\left(2 \epsilon +1,\epsilon +1,\epsilon +1,\epsilon +1,1-\epsilon ,z_1,z_2,z_3\right) \nonumber\\
	&= \sum_{i=1}^4 g_i S_i
\end{align}
where we have put the dimension $D = 2-2\e$, thus $\nu =-\e $. The $g_i$'s denote the prefactors and the Lauricella functions $F_C^{(3)}$ are denoted by $S_i$'s.

The series expansion of the prefactors $g_i$'s in $\e$ can be easily found and the expansion of the $F_C^{(3)}$ functions (i.e.,  $S_i$'s) can be obtained using the presented method. Since $\mathcal{S}$ in \eeqref{eqn:sunsetseries} is finite and the prefactors $g_i$'s can have at most second order poles, the Lauricella functions have to be expanded upto $O(\e^2)$. Thus, we have,
\begin{align}
	\mathcal{S} = \sum_{i=1}^4 g_i S_i =  \sum_{i=1}^4 \left( \sum_{j=-2}^0 \e^j g_i^{(j)} \right) \left( \sum_{k=0}^2 \e^k S_i^{(k)} \right)
\end{align}
Now we collect the coefficients of $\e$ up to $O(\e^0)$,
\begin{align}\label{eqn:S_exp}
	\mathcal{S} &= \frac{1}{\e^2}~ \mathcal{S}^{(-2)}+ \frac{1}{\e}~ \mathcal{S}^{(-1)}+ \mathcal{S}^{(0)}\nonumber\\
	&= \frac{1}{\e^2}\left[ \sum_{i=1}^4 g_i^{(-2)} S_i^{(0)} \right]+  \frac{1}{\e}\left[ \sum_{i=1}^4 \left( g_i^{(-1)} S_i^{(0)}+g_i^{(-2)} S_i^{(1)} \right) \right] + \e^0 \left[  \sum_{i=1}^4 \left( g_i^{(0)} S_i^{(0)}+ g_i^{(-1)} S_i^{(1)}+g_i^{(-2)} S_i^{(2)} \right) \right]
\end{align}

At this point, we expect that the coefficients of $\e^{-2}$ and $\e^{-1}$ in \eeqref{eqn:S_exp} to be zero as the $\mathcal{S}$ is finite. Indeed, after expanding the Lauricella functions and the prefactors in series in $\e$ and collecting the $O(\e^{-2})$ coefficient we found
\begin{align}
	\mathcal{S}^{(-2)}= \sum_{i=1}^4 g_i^{(-2)} S_i^{(0)}=0
\end{align}
However, the coefficient of $\e^{-1}$ is not identically zero, but
\begin{align}
	\mathcal{S}^{(-1)} = \sum_{i=1}^4 &\left( g_i^{(-1)} S_i^{(0)} +g_i^{(-2)} S_i^{(1)} \right) \nonumber\\
	= &-z_1  \sum_{n_1,\dots,n_4=0}^\infty \frac{(1)_{n_1} (1)_{n_4} (1)_{n_3+n_4}  (2)_{n_1+n_2+n_3+n_4}(2)_{n_1+n_2+n_3+n_4}}{ (1)_{n_2} (1)_{n_3} (2)_{n_1+n_4} (2)_{n_1+n_4}(2)_{n_3+n_4}} \frac{z_1^{n_1+n_4} z_2^{n_2} z_3^{n_3}}{n_1! n_2! n_3! n_4!}\nonumber\\
	&+z_2  \sum_{n_1,\dots,n_4=0}^\infty \frac{(1)_{n_2} (1)_{n_4} (1)_{n_3+n_4}(2)_{n_1+n_2+n_3+n_4}(2)_{n_1+n_2+n_3+n_4} }{ (1)_{n_1} (1)_{n_3} (2)_{n_2+n_4}(2)_{n_2+n_4} (2)_{n_3+n_4}} \frac{z_1^{n_1} z_2^{n_2+n_4} z_3^{n_3}}{n_1! n_2! n_3! n_4!}\nonumber\\
	&-z_2 \sum_{n_1,\dots,n_4=0}^\infty  \frac{(1)_{n_2} (1)_{n_4} (1)_{n_1+n_3+n_4}(2)_{n_1+n_2+n_3+n_4}(2)_{n_1+n_2+n_3+n_4}  }{ (1)_{n_1} (1)_{n_3} (2)_{n_2+n_4}(2)_{n_2+n_4} (2)_{n_1+n_3+n_4}} \frac{z_1^{n_1} z_2^{n_2+n_4} z_3^{n_3}}{n_1! n_2! n_3! n_4!}\nonumber\\
	&+ z_1 \sum_{n_1,\dots,n_4=0}^\infty \frac{(1)_{n_1} (1)_{n_4} (1)_{n_2+n_3+n_4} (2)_{n_1+n_2+n_3+n_4}(2)_{n_1+n_2+n_3+n_4}  }{ (1)_{n_2} (1)_{n_3} (2)_{n_1+n_4}(2)_{n_1+n_4} (2)_{n_2+n_3+n_4}} \frac{z_1^{n_1+n_4} z_2^{n_2} z_3^{n_3}}{n_1! n_2! n_3! n_4!}
\end{align}
It turns out to be zero numerically (see table \ref{table:sunset}).

We find the  coefficient of $\e^0$ (i.e., $ \mathcal{S}^{(0)}$) in a similar way but do not write it explicitly to lighten the paper. It is provided in the ancillary notebook. In Appendix \ref{sec:sunsetNum} we compare our expressions of $ \mathcal{S}^{(-2)}, \mathcal{S}^{(-1)}$ and $ \mathcal{S}^{(0)}$ against the numerical package \texttt{FIESTA5} \cite{Smirnov:2021rhf} and the $x$ space  presentation of the sunset integral in $D=2$ dimension \cite{Berends:1993ee,Groote:2012pa}, which reads
\begin{align}\label{eqn:sunsetxpace}
	T_{123}(p^2,m_1^2,m_2^2,m_3^2) = - \frac{2^4}{(2 \pi \mu^2)^{-2}}  \mathcal{I} = - \frac{2^4}{(2 \pi \mu^2)^{-2}} \int_0^\infty dx~x~ J_0( \sqrt{- p^2} x) K_0(m_1 x) K_0(m_2 x) K_0(m_3 x)
\end{align}
where $J_0$ and $K_0$ are Bessel's function of first and second kind respectively. We find that, our expression of $\mathcal{S}^{(0)}$ is in agreement with the results of \texttt{FIESTA5} and the $x$-space representation numerically.

\section{Summary}
We have presented a new method to obtain the $\e$-expansion of MHFs. In our approach, when the expansion is of Taylor type, successive derivative with respect to $\e$ has to evaluated. We have demonstrated the process of Pochhammer derivative of a MHFs which plays important role in finding the Taylor expansion. When lower Pochhammer parameters of a \hf are singular, the series expansion may be of Laurent type. In this situation, an appropriate step down operator can be found, which relates the starting \hf to another one with shifted lower Pochhammer parameters. This \hf with non-singular lower Pochhammer parameters can be expanded in Taylor series expansion. Finally the obtained step down operator can be made to act on the Taylor series of the \hf with non-singular lower Pochhammer parameters to find the $\e$-expansion of the starting function. In our method, each of the coefficient of $\e$ in the series expansion of a MHF is linear combination of \hfs with same domain of convergence as the original one.  These analytic expressions can be used to find the numerical value at given point in its argument space within the domain of convergence. The summation of such multi-fold series has to be performed with sufficient number of terms for the required precision.

This methodology allows one to find series expansion of any MHF related to the Feynman integrals. To illustrate the approach we have presented examples containing one variable Gauss $_2F_1$, two variable Appell $F_1$, $F_2$ and three variable Lauricella function, which is linked to two loop sunset Feynman integral with different masses. In the present work many of the steps of the calculations required for the examples provided in this article are performed manually using \textit{Mathematica}. We plan to build a automated  package that performs the $\e$-expansion of MHFs in near future. 

In this approach, higher order terms in the series expansion contain MHFs where all the variables are not independent. In such cases, it may be possible to reduce the fold of the hypergeometric functions. Some reduction formulae of Kamp\'e de F\'eriet series are discuss in \cite{Srivastava:1985}. The Reduction formulae for a general MHFs are unknown and beyond the scope of the paper.


\section*{Acknowledgments}
The author would like to thank B. Ananthanarayan and Tanay Pathak for the discussions and comments on the draft. We would also like to thank the Reviewer for valuable comments and suggestions which helped us in improving the quality of the manuscript. This is a part of author's doctoral work at CHEP, IISc.

	\begin{appendices}

	\section{Pochhammer identities}\label{Appendix:Pochid}
	Pochhammer parameter is ratio of Euler's gamma functions,
	\begin{align}
(a)_m =\frac{\Gamma(a+m)}{\Gamma(a)}
	\end{align}
Some useful identities of the Pochhammer parameters are listed below.
	\begin{itemize}
		\item $(a)_{m_1+\dots + m_n} = (a)_{m_1} (a+m_1)_{m_2}(a+m_1+m_2)_{m_3}\dots (a+m_1+\dots m_{n-1})_{m_n}$
		\item $(a)_{-n} \rightarrow \frac{(-1)^{n}}{(1-a)_{n}}, \quad a \notin \mathbb{Z} \text { and } n \in \mathbb{N}$
		\item Gauss's Multiplication theorem:\\ $$(a)_{k m}=k^{k m}\left(\frac{a}{k}\right)_{m}\left(\frac{a+1}{k}\right)_{m} \ldots\left(\frac{a+k-1}{k}\right)_{m} \hspace{1cm}\text{where~} k\in \mathbb{N}^*$$
	\end{itemize}

\section{Notation used for MHFs}\label{Appendix:notation}
In this article we have used a common notation of \hfs for more than one variable. We denote a \hf with $r$ fold summations with the set of  variables $\textbf{x}$ in vector notation as, 
\begin{align}
F\left[ \begin{array}{c}\left\{ \textbf{a},\textbf{m}\right\}\\
	\left\{ \textbf{b},\textbf{n}\right\}
	\end{array}\;\middle|\;
\{\textbf{x}\} \right] = \sum_{s_1=0}^{\infty}\dots\sum_{s_r=0}^{\infty} \frac{(\textbf{a})_{\textbf{m}\cdot \textbf{s}}}{(\textbf{b})_{\textbf{n}\cdot \textbf{s}}} \frac{\textbf{x}^\textbf{s}}{\textbf{s}!}
\end{align}
where $\textbf{m},\textbf{n}$ are vectors of length $p$ and $q$ respectively. Each of the elements of $\textbf{m}$ and $\textbf{n}$ themselves are vectors of length $r$  with integers in their elements.
\begin{align}
	\begin{aligned}[c]
	\textbf{s}&=\{ s_1,\dots, s_r\}\nonumber\\
	\left\{ \textbf{a},\textbf{m}\right\} &= \left\{ a_1,\mathbf{m_1}\right\}, \dots ,\left\{ a_p,\mathbf{m_p}\right\}\nonumber\\
	(\textbf{a})_{\textbf{m}\cdot \textbf{s}} &= (a_1)_{\mathbf{m_1}\cdot \textbf{s}} \dots (a_p)_{\mathbf{m_p}\cdot \textbf{s}} \nonumber\\
	\mathbf{m_j}\cdot \textbf{s} &= \sum_{i=0}^r (m_j)_i s_i \hspace{1cm} j=1,\dots,p\nonumber\\
	\textbf{x}^\textbf{s} &= \prod_{i=1}^r x_i^{s_i} 
	\end{aligned}\hspace{2cm}
\begin{aligned}[c]
\{\textbf{x}\} &= \{x_1, \dots, x_r\}\nonumber\\
\left\{ \textbf{b},\textbf{n}\right\} &= \left\{ b_1,\mathbf{n_1}\right\}, \dots ,\left\{ b_q,\mathbf{n_q}\right\}\nonumber\\
(\textbf{b})_{\textbf{n}\cdot \textbf{s}} &= (b_1)_{\mathbf{n_1}\cdot \textbf{s}} \dots (b_q)_{\mathbf{n_q}\cdot \textbf{s}} \nonumber\\
\mathbf{n_j}\cdot \textbf{s} &= \sum_{i=0}^r (n_j)_i s_i \hspace{1cm} j=1,\dots,q\nonumber\\
\textbf{x}! &= \prod_{i=1}^r (x_i!)
\end{aligned}
\end{align}
In this notation the Appell $F_1$ can be written as
\begin{align*}
	F_1 (a,b_1,b_2,c,x,y) &= \sum_{m,n=0}^\infty \frac{(a)_{m+n} (b_1)_m (b_2)_n}{(c)_{m+n}} \frac{x^m y^n}{m!n!}\nonumber\\
	&= F\left[ \begin{array}{c}\{a,\{1,1\}\}, \{b_1,\{1,0\}\}, \{b_2,\{0,1\}\}\\
		\{c,\{1,1\}\}\end{array}\;\middle|\;
	\{x,y\} \right]
\end{align*}

The Appell $F_2$ function is defined as,
\begin{align*}
	F_2\left(a, b_1, b_2 ; c_1, c_2 ; x, y\right)=\sum_{m=0}^{\infty} \sum_{n=0}^{\infty} \frac{(a)_{m+n}\left(b_1\right)_m\left(b_2\right)_n}{\left(c_1\right)_m\left(c_2\right)_n } \frac{x^m y^n}{m ! n !}
\end{align*}

\section{Numerical tests} \label{Appendix:numericaltest}
In this Appendix we test the results obtained different examples in Section \ref{sec:examples} numerically.

\subsection{Gauss $_2F_1$ example with half integer Pochhammer}\label{sec:Gausshalfintegernum}
We compare our result  from Section \ref{sec:Gausshalfinteger},(i.e. \eeqref{eqn:2f1halfinteger}) numerically with \texttt{HypExp} result (\eeqref{eqn:2f1halfintegerHypExp}) and the result obtained from \texttt{NumExp} below. The coefficients of the series expansion of $\e$ are evaluated at $x=0.3$ and tabulated in Table \ref{table:Gausshalfinteger}. We observe that our results are numerically consistent with others.

\begin{table}[H]
	\centering
	\begin{tabular}{|c|c|c|c|c|}
		\hline
		& $\e^0$ & $\e^1$    & $\e^2$   & $\e^3$    \\ \hline
		\eeqref{eqn:2f1halfinteger}        & 127.514      & -38.7907 & 610.309 & -189.456 \\ \hline
		result from \texttt{HypExp} & 127.514      & -38.7907 & 610.309 & -189.456 \\ \hline
		result from \texttt{NumExp} & 127.514      & -38.7907 & 610.309 & -189.528 \\ \hline
	\end{tabular}
	\caption{The numerical value of coefficients of $\e$ at $x=0.3$ from \eeqref{eqn:2f1halfinteger}, \texttt{HypExp} and \texttt{NumExp} are compared. The \texttt{NumExp} results are obtained with $e_h = 10^{-4}$.}\label{table:Gausshalfinteger}
\end{table}

\subsection{Gauss $_2F_1$ example} \label{sec:Gaussinteger}
We compare our result  from Section \ref{sec:Gauss}, (\eeqref{eqn:Gaussresult}) numerically with \texttt{HypExp} result (\eeqref{eqn:GaussresultHypExp}) and the result obtained from \texttt{NumExp} below. The coefficients of the series expansion of $\e$ are evaluated at $x=0.1$ and tabulated in Table \ref{table:Gauss}. We observe that our results are numerically consistent with others.

\begin{table}[H]
	\centering
	\begin{tabular}{|c|c|c|c|c|}
		\hline
		& $\e^0$ & $\e^1$    & $\e^2$   & $\e^3$    \\ \hline
		\eeqref{eqn:Gaussresult}        & 1      & 0.0057506 & 0.104955 & 0.0998262 \\ \hline
		result from \texttt{HypExp} & 1      & 0.0057506 & 0.104955 & 0.0998262 \\ \hline
		result from \texttt{NumExp} & 1      & 0.0057506 & 0.104955 & 0.099735  \\ \hline
	\end{tabular}
\caption{The numerical value of coefficients of $\e$ at $x=0.1$ from \eeqref{eqn:Gaussresult}, \texttt{HypExp} and \texttt{NumExp} are compared.  The \texttt{NumExp} results are obtained with $e_h = 10^{-4}$.}\label{table:Gauss}
\end{table}

 \subsection{Taylor series of Appell $F_1$}\label{sec:PhanF1}
 In this section we numerically evaluate the $\e$-expansion coefficients of the Appell $F_1$ function (\eeqref{eqn:PhanAppellF1exp}) at $(x_c,y_c) = (11.1-10^{-12} i, 12.1 -10^{-12} i)$. The obtained numerical values are compared with the result given in \cite{Phan:2018cnz} in Table \ref{table:PhanF1} below.

 \begin{table}[!htbp]
 	\centering
 	\begin{tabular}{|c|c|}
 		\hline
 		 Numerical value of expansion coefficients of \eeqref{eqn:PhanAppellF1exp} & Results from Table B.2 of \cite{Phan:2018cnz}  \\ \hline
 		 (-0.17504424807351878429 - 0.05422812947330403050 i )$\e^0$ & 
 		(-0.17504424807358806571 -0.05422812947328981004i)$\e^0$ \\ \hline
 		(-0.00861885859121239314 - 0.39051763820460791405 I)$\e^1$ & (-0.00861885859131501092 -0.39051763820462137566i)$\e^1$ \\ \hline
 		(0.37518853545322500566 - 0.34047477405513330909 I)$\e^2$& (0.37518853545319785781 -0.34047477405516524129i)$\e^2$ \\ \hline
 		(0.49765461883465610385 - 0.00717399489426463939 I)$\e^3$ & (0.49765461883470790694 -0.00717399489427550385i)$\e^3$\\ \hline
 		(0.32835724868568101641 + 0.23005850008096157264 I)$\e^4$& (0.32835724868237320395 + 0.23005850008124251183i)$\e^4$\\ \hline
 	\end{tabular}
 	\caption{We compare the numerical results of the coefficients of the series expansion of the Appell function (\eeqref{eqn:PhanAppellF1exp}) upto $\e^4$ with the results provided in Table B.2 of \cite{Phan:2018cnz}. The values used for the numerical evaluation is $(x_c,y_c) = (11.1-10^{-12} i, 12.1 -10^{-12} i)$. The upper limit of the sum is taken as $\text{N}=40$ for each of the summation indices while evaluating the coefficients of $\e^0,\e^1$ and $\e^2$. For the coefficients of $\e^3$ and $\e^4$,  N is taken as $20$ and $10$ respectively. To obtain the numerical value with more accuracy, one may take larger values of N, but the time taken for numerical evaluation of higher order terms increases rapidly due to presence of higher fold series.}\label{table:PhanF1}
 \end{table}

\subsection{Laurent series of Appell $F_1$}\label{sec:AppellF1num}
To test our result numerically, we multiply $\e$ to the both sides of the Laurent expansion of the Appell function considered in Section \ref{sec:AppellF1example}.
The right hand side becomes a series with non-negative powers of $\e$.
\begin{align*}
	\e ~F_1(3/2 , 2\e +1, 4-\e, \e-2,x,y) = C_0 + C_1 \e + C_2 \e^2 + O(\e^3)
\end{align*}

Each of the coefficients with appropriate powers of $\e$ is evaluated at $(x,y)=(0.1,0.2)$ with $\e = 10^{-3}$ with sufficient numbers of terms and tabulated in  Table \ref{table:AppellF1part1}

\begin{table}[!htbp]
	\centering
	\begin{tabular}{|c|c|c|}
			\hline
			$C_0$ & $ C_1 \e$ & $C_2 \e^2$  \\ \hline
			8.511583950834586989 & 0.00180071895404555128 & 
			0.00001320237872566561435 \\ \hline
		\end{tabular}
	\caption{The values of $C_i \e^i$, for $i = 0,1,2$ at $x=0.1$ and $ y=0.2$ with $\e=10^{-3}$ are shown.}\label{table:AppellF1part1}
\end{table}
The numerical value of $ \e ~F_1(3/2 , 2\e +1, 4-\e, \e-2,x,y) $ is calculated with same $x,y$ and $\e$ and  compared with sum the first three terms of its series expansion in Table \ref{table:AppellF1part2}.

\begin{table}[!htbp]
	\centering
	\begin{tabular}{|c|c|}
		\hline
		$\e ~F_1(3/2 , 2\e +1, 4-\e, \e-2,x,y) $ & $C_0 + C_1 \e + C_2 \e^2 $   \\ \hline
		8.513397873773039908   & 8.51339787216735821 \\ \hline
	\end{tabular}
	\caption{Comparison of the values of both sides of the series of $\e ~F_1(3/2 , 2\e +1, 4-\e, \e-2,x,y) $  upto $\e^2$ terms at $(x,y)=(0.1,0.2)$ with $\e=10^{-3}$. The value of the series expansion upto $\e^2$ terms is in agreement upto 8 decimal places with value of $\e ~F_1(\dots,x,y)$. }\label{table:AppellF1part2}
\end{table}

\subsection{Appell $F_2$ example} \label{sec:MochF2num}

In this section we numerically compute the $\e$-expansion coefficients of the Appell $F_2$ function (\eeqref{eqn:genAppellF2exp}) at $(x,y) = (0.6,0.1)$ with the $a_i$'s equal to $1$. The obtained numerical values are compared with the numerical value of the Eq. (81) of \cite{Moch:2001zr} in Table \ref{table:MochF1} below.

\begin{table}[!htbp]
	\centering
	\begin{tabular}{|c|c|c|c|}
		\hline
	Order& N &	Numerical value of expansion  & Numerical value of expansion   \\ 
	&	&coefficients of \eeqref{eqn:genAppellF2exp} with $a_i=1$& coefficients of Eq. (81) of \cite{Moch:2001zr} \\\hline
 		& 25 &2.4999957354567953718  & 
		2.4999999999999998612\\ 
		$\e^0$& 50 & 2.4999999999878756421 & \\
		& 75 & 2.4999999999999998268 & \\
		& 100 & 2.4999999999999998612 & \\\hline
		 & 25 & -1.571552060080885826 &-1.571521648555935118\\
		$\e^1$& 50 & -1.571521649284227887 & \\
		& 75 & -1.571521648555953585 & \\
		& 100 & -1.571521648555935119 & \\\hline
		& 25 & 2.608431603015495386 & 2.608484299442508259 \\ 
		$\e^2$ & 50 & 2.608484297358779548 & \\
		& 75 & 2.608484299442436849 & \\
		& 100 & 2.608484299442508257 & \\\hline
		 & 25 & -2.154453137514624822 &-2.154366134071260766\\
		$\e^3$& 50 & -2.154366138151998583 & \\
		& 75 & -2.154366134071422536 & \\\hline
	\end{tabular}
	\caption{We compare the numerical results of the coefficients of the series expansion of the Appell function (\eeqref{eqn:genAppellF2exp}) with all the  $a_i=1$ up to $\e^3$ with the results provided in Eq. (81) of \cite{Moch:2001zr}. The values used for the numerical evaluation is $(x,y) = (0.6,0.1)$. Here, as always, N is the upper limit of the summation for each of the indices. We evaluate the series coefficients with a set to values of N, to show how the values changes with the upper limit of the summation. The numerical results are found to be in good agreement with each other. To obtain the numerical value with more accuracy, one may take larger values of N, but the time taken for numerical evaluation of higher order terms increases rapidly due to presence of higher fold series.  }\label{table:MochF1}
\end{table}

\vfill

\subsection{Appell $F_2$ example with integers, half-integers and rational parameters}
\label{sec:parameterindependence}

In this section we continue the same example of Appell $F_2$ (i.e., \eeqref{eqn:genAppellF2exp}) and perform numerical tests where $a_i$'s take values from integers, half-integers and rational numbers. Each of the series coefficients is evaluated with a set of upper limit (N) to show the dependencies of the results on N.

First, we consider the case where $a_i$'s take integer values. Writing the series expansion of $F_2$ with the following values of $a_i$'s :  $a_1 = 1, a_2 = 2, a_3 = 3$ and $a_4 = 4$
\begin{align} \label{eqn:F2int}
	F_2 (1, 2 ,\e , 3+\e, 4-\e,x,y)  = f_0 + \e  f_1 + \e^2 f_2 + \e^3 f_3
\end{align} 
 We find the numerical values of each of the series coefficients of \eeqref{eqn:F2int} with the upper limit $N = 15,30,45$ and $60$ at $(x,y) = (0.2,0.3)$ in Table \ref{table:F2int} below.

\begin{table}[H]
	\centering
	\begin{tabular}{|c|c|c|c|c|}
		\hline
		N & $f_0$ &	$f_1$  & $f_2$ & $f_3$  \\ \hline
		15 & 1.15717756570958926382 &	0.04946518465296974939  & 0.04760586647909068011 & 0.00446404842334520315\\
		30 & 1.15717756571048779862 & 0.04946518465764881786	  & 0.04760586648282076076 & 0.00446404842749894999\\
		45 & 1.15717756571048779862 &	0.04946518465764881791  & 0.04760586648282076074  & 0.00446404842749895002\\
		60 & 1.15717756571048779862 &	0.04946518465764881791  & 0.04760586648282076074  & 0.00446404842749895002 \\\hline
	\end{tabular}
	\caption{We compute the first four series coefficients of expansion of $F_2 (1, 2 ,\e , 3+\e, 4-\e,x,y)$ at $(x,y) = (0.2,0.3)$. The summations are performed with upper limit N for each of the summation indices.}\label{table:F2int}
\end{table}

Now we compare both sides of \eeqref{eqn:F2int} with $\e = 10^{-5}$ in Table \ref{table:F2intnum}.
\begin{table}[H]
	\centering
	\begin{tabular}{|c|c|c|}
		\hline
	  & & $F_2 (1, 2 ,\e , 3+\e, 4-\e,x,y)$  \\ \hline
	$f_0$ & 1.15717756571048779862 & \\
	$f_0 + \e  f_1  $ & 1.15717806036233437511 &\\
	$f_0 + \e  f_1 + \e^2 f_2 $& 1.15717806036709496176 &\\
$f_0 + \e  f_1 + \e^2 f_2 + \e^3 f_3 $	& 1.15717806036709496622 & 1.15717806036709496622\\ \hline
	\end{tabular}
	\caption{We compare both sides of the series  expansion of $F_2 (1, 2 ,\e , 3+\e, 4-\e,x,y)$ (i.e., \eqref{eqn:F2int}) at $(x,y) = (0.2,0.3)$. The summations are performed with the upper limit $\text{N}=60$ for each of the  indices. Both the sides of \eeqref{eqn:F2int} are found to be in good agreement.}\label{table:F2intnum}
\end{table}

Next, we take half-integer values of $a_i$'s :  $a_1 = 1/2, a_2 = 3/2, a_3 = 5/2$ and $a_4 = 7/2$
\begin{align} \label{eqn:F2halfint}
	F_2 (\frac{1}{2},\frac{3}{2},\epsilon ,\epsilon +\frac{5}{2},\frac{7}{2}-\epsilon,x,y)  = g_0 + \e  g_1 + \e^2 g_2 + \e^3 g_3
\end{align} 

Similar to the previous case, we compute the series coefficients of \eeqref{eqn:F2halfint} at $(x,y) = (0.2,0.3)$ with the upper limit $\text{N} = 15,30,45$ and $60$ in Table \ref{table:F2halfint} below.

\begin{table}[H]
	\centering
	\begin{tabular}{|c|c|c|c|c|}
		\hline
		N & $g_0$ &	$g_1$  & $g_2$ & $g_3$  \\ \hline
		15 & 1.06740285248337825566 & 0.02629391741076840821 & 0.02793237719286772142 & 0.00201506616058023520 \\
		30 & 1.06740285248347457546 & 0.02629391741162192975	  & 0.02793237719337559810 & 0.00201506616133570611\\
		45 & 1.06740285248347457546 & 0.02629391741162192976 & 0.02793237719337559810  & 0.00201506616133570612\\
		60 & 1.06740285248347457546 &	0.02629391741162192976  & 0.02793237719337559810  & 0.00201506616133570612 \\\hline
	\end{tabular}
	\caption{We compute the first four series coefficients of expansion of $F_2 (1/2, 3/2 ,\e , 5/2+\e, 7/2-\e,x,y)$ at $(x,y) = (0.2,0.3)$ with upper limit $N$ for each of the summation indices.}\label{table:F2halfint}
\end{table}

We compare both sides of \eqref{eqn:F2halfint} with $\e = 10^{-5}$ below in Table \ref{table:F2halfintnum}

\begin{table}[H]
	\centering
	\begin{tabular}{|c|c|c|}
		\hline
		& & $F_2 (1/2, 3/2 ,\e , 5/2+\e, 7/2-\e,x,y)$  \\ \hline
		$g_0$ & 1.06740285248347457546 & \\
		$g_0 + \e  g_1  $ & 1.06740311542264869168 &\\
		$g_0 + \e  g_1 + \e^2 g_2 $& 1.06740311542544192940 &\\
		$g_0 + \e  g_1 + \e^2 g_2 + \e^3 g_3 $	& 1.06740311542544193141 & 1.06740311542544193141\\ \hline
		
	\end{tabular}
	\caption{We compare both sides of the series  expansion of $F_2 (1/2, 3/2 ,\e , 5/2+\e, 7/2-\e,x,y)$ (i.e., \eqref{eqn:F2halfint}) at $(x,y) = (0.2,0.3)$. The summations are performed with the upper limit $\text{N}=60$ for each of the  indices. Both the sides of \eeqref{eqn:F2halfint} are found to be in good agreement.}\label{table:F2halfintnum}
\end{table}

Lastly, we choose rational values for $a_i$'s, $a_1 = 1/7, a_2 = 2/7, a_3 = 3/7$ and $a_4 = 4/7$
\begin{align} \label{eqn:F2rational}
	F_2 (\frac{1}{7},\frac{2}{7},\epsilon ,\epsilon +\frac{3}{7},\frac{4}{7}-\epsilon,x,y)  = h_0 + \e  h_1 + \e^2 h_2 + \e^3 h_3
\end{align} 
We now compute the series coefficients of \eeqref{eqn:F2rational} at $(x,y) = (0.2,0.3)$ with the upper limit $\text{N} = 15,30,45$ and $60$ in Table \ref{table:F2rational}.

\begin{table}[H]
	\centering
	\begin{tabular}{|c|c|c|c|c|}
		\hline
		N & $h_0$ &	$h_1$  & $h_2$ & $h_3$  \\ \hline
		15 & 1.0213175368701332097 & 0.052073040471508997983 & 0.28243204145121175747 & 0.10874010806468834360 \\
		30 & 1.0213175368701836808  & 0.052073040691456832163	  & 0.28243204129337127545 & 0.10874010942130990672\\
		45 & 1.0213175368701836808  & 0.052073040691456863666 & 0.28243204129337121899 &0.10874010942131012250 \\
		60 & 1.0213175368701836808 & 0.052073040691456863666	  & 0.28243204129337121899  & 0.10874010942131012250 \\\hline
	\end{tabular}
	\caption{We compute the first four series coefficients of expansion of $F_2 (1/7, 2/7 ,\e , 3/7+\e, 4/7-\e,x,y)$ at $(x,y) = (0.2,0.3)$ with upper limit N for each of the summation indices.}\label{table:F2rational}
\end{table}

Now we compare both sides of \eqref{eqn:F2rational} with $\e = 10^{-5}$ in Table \ref{table:F2numrational}.

\begin{table}[H]
	\centering
	\begin{tabular}{|c|c|c|}
		\hline
		& & $F_2 (1/7, 2/7 ,\e , 3/7+\e, 4/7-\e,x,y)$  \\ \hline
		$h_0$ & 1.0213175368701836808 & \\
		$h_0 + \e  h_1  $ & 1.0213180576005905954 &\\
		$h_0 + \e  h_1 + \e^2 h_2 $& 1.0213180576288337995 &\\
		$h_0 + \e  h_1 + \e^2 h_2 + \e^3 h_3 $	& 1.0213180576288339083 & 1.0213180576288339083\\ \hline
		
	\end{tabular}
	\caption{We compare both sides of the series  expansion of $F_2 (1/7, 2/7 ,\e , 3/7+\e, 4/7-\e,x,y)$ (i.e., \eqref{eqn:F2rational}) at $(x,y) = (0.2,0.3)$. The summations are performed with the upper limit $\text{N}=60$ for each of the  indices. Both the sides of \eeqref{eqn:F2rational} are found to be in good agreement.}\label{table:F2numrational}
\end{table}

We observe that the coefficients of the series expansion of $F_2$ in \eeqref{eqn:genAppellF2exp} convergence in a same rate, independent of the values of the parameters. In all the cases, at $(x,y)=(0.2,0.3)$, the  upper limit of the summation $\text{N}=60$ of the series coefficients is enough to obtain a match in numerical results with the LHS up to 20 digits.

\subsection{The Sunset example} \label{sec:sunsetNum}

In this section we numerically compare our results of $\mathcal{S}^{(-2)}, \mathcal{S}^{(-1)}$ and $\mathcal{S}^{(0)}$ with the $x$-space representation of the sunset integral and the result obtained by the numerical package \texttt{FIESTA5} \cite{Smirnov:2021rhf}. We recall that, $\mathcal{S}$, given by \eeqref{eqn:sunsetseries}, is expanded in Laurent series in $\e (= (2-D)/2)$
\begin{align*}
	\mathcal{S} =\frac{1}{\e^2}~ \mathcal{S}^{(-2)}+ \frac{1}{\e}~ \mathcal{S}^{(-1)}+ \mathcal{S}^{(0)}
\end{align*}
Since the sunset integral is finite in $D=2$ dimension, we expect $\mathcal{S}^{(-2)}$ and $ \mathcal{S}^{(-1)}$ to be zero.  Thus we compare $\mathcal{S}^{(0)}$ with the $x$-space representation, given by  \eeqref{eqn:sunsetxpace}. Fixing the overall prefactors, we expect the two quantities below to be equal.
\begin{align}
	\mathcal{S}^{(0)} \stackrel{?}{=} 4 m_3^2~ \mathcal{I}
\end{align}
where the integral $\mathcal{I}$, given by \eeqref{eqn:sunsetxpace}, is calculated in \textit{Mathematica} using the \texttt{NIntegrate} command.

The coefficients of the series expansion of $\mathcal{S}$ (i.e., $\mathcal{S}^{(-2)}, \mathcal{S}^{(-1)}$ and $\mathcal{S}^{(0)}$) are calculated for three different kinematic points (within the domain of convergence of the Lauricella series) specified by a set of values of $p^2, m_1^2, m_2^2$ and $m_3^2$ and tabulated in the Table  \ref{table:sunset}. For each set, the coefficient of $\e^{-2}$ (i.e., $\mathcal{S}^{(-2)}$) is identically equal to zero. The coefficients of $\e^{-1}$ and $\e^0$ (i.e., $\mathcal{S}^{(-1)}$ and $\mathcal{S}^{(0)}$ ) are calculated with different values of  the upper limit N for each of the summation indices to show the dependence of the result on N. For example, in the first kinematic point, given by $(p^2,m_1^2, m_2^2, m_3^2) = (3,10,2,100)$, $\mathcal{S}^{(-1)}$ is calculated with $\text{N}= 10,15,20,25$.  As the number of terms in the summation is increased, the value of $\mathcal{S}^{(-1)}$ approaches to zero. Similar trend follows for the second and third kinematic points.  $\mathcal{S}^{(0)}$ is calculated with $\text{N}=10,15,20$ for the first kinematic point. The result with the upper limit of the summation N = 20 matches with the result of the $x$-space representation (i.e., $4 m_3^2~ \mathcal{I}$). The numerical results of the sunset integral from \texttt{FIESTA5} are also provided for comparison.

\begin{table}[!htbp]
	\begin{tabular}{|c|c|c|c|c|c|c|c|}
		\hline
		kinematic &
		$\mathcal{S}^{(-2)}$ &
		$\mathcal{S}^{(-1)}$ &
		&
		$\mathcal{S}^{(0)}$ &
		&
		result from &
		result from  the\\ \cline{2-6}
		points &   &   & N &   & N & \texttt{FIESTA5} & integration ( $4 m_3^2~ \mathcal{I}$ )\\ \hline
		&   &   &   &   &   &        &             \\
		$p^2$ = 3  & 0 & $-1.09434200450461*10^{-9}$ & 10 & 13.0801528975881 & 10 & 13.08015  & 13.0801528947195         \\
		$m_1^2$ = 10             &  & $-3.46337421335846*10^{-13}$ & 15 & 13.0801528947215 & 15 & $\pm 0.00007$       &            \\
		$m_2^2$ = 2               &  & $-1.18293156085287*10^{-16}$ & 20 & 13.0801528947195 & 20 &       &            \\
		$m_3^2$ = 100              &  & $-4.24195869926179*10^{-20}$ & 25 &  &  &       &            \\
		&   &   &   &   &   &        &             \\ \hline
		&   &   &   &   &   &        &             \\
		$p^2$ = 3  & 0 & $4.60503575832032*10^{-18}$ & 10 & 21.8600302323743 & 10 & 21.8600  & 21.8600302323744         \\
		$m_1^2$ = 1             &  & $2.64103442427579*10^{-25}$ & 15 & 21.8600302323744 & 15 & $\pm 0.0001$       &            \\
		$m_2^2$ = 2               &  & $1.60726507054367*10^{-32}$ & 20 &  &  &       &            \\
		$m_3^2$ = 100              &  & $1.02402152695267*10^{-39}$ & 25 &  &  &       &            \\
		&   &   &   &   &   &        &             \\ \hline
		&   &   &   &   &   &        &             \\
		$p^2$ = 3  & 0 & $3.62652142904041*10^{-31}$ & 10 & 46.3800201955786 & 10 & 46.3800 & 46.3800201955786         \\
		$m_1^2$ = 1             &  & $3.13801126225243*10^{-44}$ & 15 &  &  & $\pm 0.0003$       &            \\
		$m_2^2$ = 2               &  & $2.41027454208700*10^{-57}$ & 20 & &  &       &            \\
		$m_3^2$ = 1000              &  &  &  &  &  &       &            \\
		&   &   &   &   &   &        &             \\ \hline
	\end{tabular}
\caption{We calculate $\mathcal{S}^{(-2)}, \mathcal{S}^{(-1)}$ and $\mathcal{S}^{(0)}$ for three different kinematic points and compare the results with the results obtained from \texttt{FIESTA5} and by numerically performing the integration in \eqref{eqn:sunsetxpace} using \textit{Mathematica}. }
\label{table:sunset}
\end{table}

	\vfill

\section{The $_2F_2$ example}
In this example we apply the methodology to confluent hypergeometric function  $_2F_2$. Consider the following function
\begin{align} \label{eqn:2f2}
	G \vcentcolon= ~ _2F_2(1,2;\e,\e-1,x) = \sum_{m=0}^\infty \frac{(1)_m (2)_m }{ (\e)_m (\e-1)_m} \frac{x^m}{m!} \hspace{1cm}\text{where~} |x|<\infty
\end{align}
Note that the lower Pochhammer parameters are $\e$ and $\e-1$. So, the function $G$ may have Laurent series expansion. On the other hand, the function $_2F_2(1,2;\e+1,\e+1,x)$ has Taylor series expansion. We find the step down operator that relates to the former i.e. \eeqref{eqn:2f2} with the later
\begin{align}\label{eqn:2f2stepdown}
	G = H\bullet ~_2F_2(1,2;\e+1,\e+1,x)
\end{align}
The step down operator is found to be
\begin{align}
	H= \frac{\e^3-\e^2+2 x}{(\e-1) \e^2} +\frac{x  \left(2 \e^2-\e+4 x-1\right)}{(\e-1) \e^2} \partial_x +\frac{x^2  (\e+x-1)}{(\e-1) \e^2} \partial_x^2
\end{align}
Expanding the operator $H$ in series, and collecting the coefficients of $\e$ we find
\begin{align}\label{eqn:2f2stepdownseries}
	H &= \left[-(x-1) x^2 \partial_x^2 + x (1-4 x) \partial_x-2 x\right]\frac{1}{\e^2} + \left[-x^3 \partial_x^2+(2-4 x) x \partial_x-2 x\right] \frac{1}{\e}\nonumber\\
	&+ \left[- x^3 \partial_x^2-4 x^2 \partial_x-2 x+1\right] + \left[ - x^3 \partial_x^2-4  x^2 \partial_x-2  x \right] \e + O(\e^2)
\end{align}
The Taylor expansion of the $_2F_2$ function on the right and side of \eeqref{eqn:2f2stepdown} is found to be
\begin{align} \label{eqn:2f2Taylor}
	~_2F_2(1,2&;\e+1,\e+1,x) = ~_1F_1(2;1,x) +    -4 \e x~F\left[ \begin{array}{c}\{3,\{1,1\}\}, \{1,\{1,0\}\}, \{1,\{0,1\}\}, \{1,\{0,1\}\}\\
		\{2,\{1,1\}\}, \{2,\{1,1\}\}, \{2,\{0,1\}\}\end{array}\;\middle|\;
	\{x,x\} \right]\nonumber\\
	&+ 6 x \e^2 ~F\left[ \begin{array}{c}\{3,\{1,1\}\}, \{1,\{1,0\}\}, \{1,\{0,1\}\}, \{1,\{0,1\}\}\\
		\{2,\{1,1\}\}, \{2,\{1,1\}\}, \{2,\{0,1\}\}\end{array}\;\middle|\;
	\{x,x\} \right]\nonumber\\
	&+ \frac{3 x^2}{4} \e^2 ~F\left[ \begin{array}{c}\{4,\{1,1,1\}\}, \{1,\{1,0,0\}\}, \{2,\{0,1,1\}\}, \{1,\{0,1,0\}\}, \{1,\{0,1,0\}\}, \{1,\{0,0,1\}\}\\
		\{3,\{1,1,1\}\}, \{3,\{1,1,1\}\}, \{3,\{0,1,1\}\}, \{2,\{0,1,0\}\}\end{array}\;\middle|\;
	\{x,x,x\} \right]\nonumber\\
	&+ \frac{9 x^2}{8} \e^2 ~F\left[ \begin{array}{c}\{4,\{1,1,1\}\}, \{1,\{1,0,0\}\}, \{2,\{0,1,1\}\}, \{1,\{0,1,0\}\}, \{2,\{0,0,1\}\}, \{1,\{0,0,1\}\}\\
		\{3,\{1,1,1\}\}, \{3,\{1,1,1\}\}, \{3,\{0,1,1\}\}, \{3,\{0,0,1\}\}\end{array}\;\middle|\;
	\{x,x,x\} \right]\nonumber\\
	&+ O(\e^3)
\end{align}
To find the Laurent series expansion of \eeqref{eqn:2f2}, we apply the step down operator \eeqref{eqn:2f2stepdownseries} on the Taylor series expansion of $_2F_2(1,2;\e+1,\e+1,x)$, i.e. \eeqref{eqn:2f2Taylor}. The Laurent  expansion reads,
\begin{align}\label{eqn:2f2result}
	G &= \left[-2 x \, _1F_1(2;1;x) + 2 (1-4 x) x \, _1F_1(3;2;x) -3 (x-1) x^2 \, _1F_1(4;3;x)\right]\frac{1}{\e^2}\nonumber\\
	& + \Bigg[-2 x \, _1F_1(2;1;x) + 2 (2-4 x) x \, _1F_1(3;2;x) -3 x^3 \, _1F_1(4;3;x) \nonumber\\
	&+ 4 x (6 x-1)~F\left[ \begin{array}{c}\{3,\{1,1\}\}, \{1,\{1,0\}\}, \{1,\{0,1\}\}, \{1,\{0,1\}\}\\
		\{2,\{1,1\}\}, \{2,\{1,1\}\}, \{2,\{0,1\}\}\end{array}\;\middle|\;
	\{x,x\} \right]\nonumber\\
	&+9 x^2 (2 x-1)~F\left[ \begin{array}{c}\{4,\{1,1\}\}, \{2,\{1,0\}\}, \{1,\{0,1\}\}, \{1,\{0,1\}\}\\
		\{3,\{1,1\}\}, \{3,\{1,1\}\}, \{2,\{0,1\}\}\end{array}\;\middle|\;
	\{x,x\} \right]\nonumber\\
	&+\frac{9 x^2 (2 x-1)}{2 }~F\left[ \begin{array}{c}\{4,\{1,1\}\}, \{1,\{1,0\}\}, \{2,\{0,1\}\}, \{2,\{0,1\}\}\\
		\{3,\{1,1\}\}, \{3,\{1,1\}\}, \{3,\{0,1\}\}\end{array}\;\middle|\;
	\{x,x\} \right]\nonumber\\
	&+\frac{8 (x-1) x^3}{3 }~F\left[ \begin{array}{c}\{5,\{1,1\}\}, \{3,\{1,0\}\}, \{1,\{0,1\}\}, \{1,\{0,1\}\}\\
		\{4,\{1,1\}\}, \{4,\{1,1\}\}, \{2,\{0,1\}\}\end{array}\;\middle|\;
	\{x,x\} \right]\nonumber\\
	&+\frac{4 (x-1) x^3}{3}~F\left[ \begin{array}{c}\{5,\{1,1\}\}, \{2,\{1,0\}\}, \{2,\{0,1\}\}, \{2,\{0,1\}\}\\
		\{4,\{1,1\}\}, \{4,\{1,1\}\}, \{3,\{0,1\}\}\end{array}\;\middle|\;
	\{x,x\} \right]\nonumber\\
	&+ \frac{8 (x-1) x^3}{9 }~F\left[ \begin{array}{c}\{5,\{1,1\}\}, \{1,\{1,0\}\}, \{3,\{0,1\}\}, \{3,\{0,1\}\}\\
		\{4,\{1,1\}\}, \{4,\{1,1\}\}, \{4,\{0,1\}\}\end{array}\;\middle|\;
	\{x,x\} \right]
	\Bigg]\frac{1}{\e} + O(\e^0)
\end{align}

The next two higher order terms are also evaluated and can be found in the ancillary file. A numerical test of the above result can be found in Appendix  \ref{sec:2f2numericaltest}.


\subsection{The $_2F_2$ example}\label{sec:2f2numericaltest}
To check the result (i.e. \eeqref{eqn:2f2result}) of the $_2F_2$ example, we multiply both sides of \eeqref{eqn:2f2result} by $\e^2$. The right hand side becomes a series with non-negative powers of $\e$.
\begin{align*}
	\e^2 ~_2F_2(1,2;\e,\e-1,x) = C_0 + C_1 \e + C_2 \e^2 + C_3 \e^3+ O(\e^4)
\end{align*}
Each of the coefficients with appropriate powers of $\e$  are evaluated at $x=0.1$ with $\e=10^{-4}$ and tabulated in Table \ref{table:2f2part1}.

\begin{table}[!htbp]
	\centering
	\begin{tabular}{|c|c|}
		\hline
		$C_0$ & -0.07305179768480031690\\ \hline
		$C_1 \e$ & -1.795147633233522925*10$^{-5}$\\ \hline
		$C_2 \e^2$ & 7.021511837197325735*10$^{-9}$\\ \hline
		$C_3 \e^3$ &  -1.52211510291782210*10$^{-13}$\\
		\hline
	\end{tabular}
	\caption{The values of $C_i \e^i$, for $i = 0,1,2,3$ at $x=0.1$ with $\e=10^{-4}$ are shown.}\label{table:2f2part1}
\end{table}

The left hand side i.e. $\e^2~ _2F_2(1,2;\e,\e-1,x)$ is calculated at the same $x$ and $\e$ with sufficient number of terms and compared with the sum of $C_0, C_1 \e , C_2 \e^2 $ and $ C_3 \e^3$ in the Table \ref{table:2f2part2}.

\begin{table}[!htbp]
	\centering
	\begin{tabular}{|c|c|}
		\hline
		$\e^2 ~_2F_2(1,2;\e,\e-1,x) $ & $C_0 + C_1 \e + C_2 \e^2 + C_3 \e^3$   \\ \hline
		-0.07306974213977305910   & -0.0730697421397730264 \\ \hline
	\end{tabular}
	\caption{Comparison of the values of both sides of the series of $\e^2~ _2F_2(1,2;\e,\e-1,x)$  upto $\e^3$ terms at $x=0.1$ with $\e=10^{-4}$. The result from the series expansion upto $\e^3$ terms is in agreement with the value of $\e^2 ~_2F_2(\dots,x) $  }\label{table:2f2part2}
\end{table}

	\end{appendices}
	\pagebreak
	\printbibliography
	
\end{document}